\documentclass[12pt,a4paper]{article}
\usepackage{xcolor,lineno}
\usepackage[pdftex,colorlinks=true]{hyperref}
\definecolor{darkblue}{rgb}{0,0,.6}
\hypersetup{citecolor=darkblue,linkcolor=darkblue,urlcolor=darkblue}

\usepackage{amsmath,amssymb,amsfonts,amsthm,mathtools,bm}
\usepackage[english]{babel}
\usepackage{makecell,multirow,orcidlink}
\AtBeginDocument{}

\usepackage{iftex}
\ifPDFTeX
  \usepackage[T1]{fontenc}
  \usepackage[utf8]{inputenc}
  \usepackage{textcomp} 
\else 
  \usepackage{unicode-math}
  \defaultfontfeatures{Scale=MatchLowercase}
  \defaultfontfeatures[\rmfamily]{Ligatures=TeX,Scale=1}
\fi
\usepackage{lmodern}
\ifPDFTeX\else  
\fi
\IfFileExists{upquote.sty}{\usepackage{upquote}}{}
\IfFileExists{microtype.sty}{
  \usepackage[]{microtype}
  \UseMicrotypeSet[protrusion]{basicmath} 
}{}
\makeatletter
\@ifundefined{KOMAClassName}{
  \IfFileExists{parskip.sty}{%
    \usepackage{parskip}
  }{
    \setlength{\parindent}{0pt}
    \setlength{\parskip}{6pt plus 2pt minus 1pt}}
}{
  \KOMAoptions{parskip=half}}
\makeatother
\makeatletter
\ifx\paragraph\undefined\else
  \let\oldparagraph\paragraph
  \renewcommand{\paragraph}{
    \@ifstar
      \xxxParagraphStar
      \xxxParagraphNoStar
  }
  \newcommand{\xxxParagraphStar}[1]{\oldparagraph*{#1}\mbox{}}
  \newcommand{\xxxParagraphNoStar}[1]{\oldparagraph{#1}\mbox{}}
\fi
\ifx\subparagraph\undefined\else
  \let\oldsubparagraph\subparagraph
  \renewcommand{\subparagraph}{
    \@ifstar
      \xxxSubParagraphStar
      \xxxSubParagraphNoStar
  }
  \newcommand{\xxxSubParagraphStar}[1]{\oldsubparagraph*{#1}\mbox{}}
  \newcommand{\xxxSubParagraphNoStar}[1]{\oldsubparagraph{#1}\mbox{}}
\fi
\makeatother

\usepackage{longtable,booktabs,array}
\usepackage{calc} 
\usepackage{etoolbox}
\makeatletter
\patchcmd\longtable{\par}{\if@noskipsec\mbox{}\fi\par}{}{}
\makeatother
\IfFileExists{footnotehyper.sty}{\usepackage{footnotehyper}}{\usepackage{footnote}}
\makesavenoteenv{longtable}
\usepackage{graphicx}
\makeatletter
\def\maxwidth{\ifdim\Gin@nat@width>\linewidth\linewidth\else\Gin@nat@width\fi}
\def\maxheight{\ifdim\Gin@nat@height>\textheight\textheight\else\Gin@nat@height\fi}
\makeatother
\setkeys{Gin}{width=\maxwidth,height=\maxheight,keepaspectratio}
\makeatletter
\def\fps@figure{htbp}
\makeatother

\makeatletter
\@ifpackageloaded{caption}{}{\usepackage{caption}}
\AtBeginDocument{%
\ifdefined\contentsname
  \renewcommand*\contentsname{Table of contents}
\else
  \newcommand\contentsname{Table of contents}
\fi
\ifdefined\listfigurename
  \renewcommand*\listfigurename{List of Figures}
\else
  \newcommand\listfigurename{List of Figures}
\fi
\ifdefined\listtablename
  \renewcommand*\listtablename{List of Tables}
\else
  \newcommand\listtablename{List of Tables}
\fi
\ifdefined\figurename
  \renewcommand*\figurename{Figure}
\else
  \newcommand\figurename{Figure}
\fi
\ifdefined\tablename
  \renewcommand*\tablename{Table}
\else
  \newcommand\tablename{Table}
\fi
}
\@ifpackageloaded{float}{}{\usepackage{float}}
\floatstyle{ruled}
\@ifundefined{c@chapter}{\newfloat{codelisting}{h}{lop}}{\newfloat{codelisting}{h}{lop}[chapter]}
\floatname{codelisting}{Listing}

\makeatother
\usepackage{setspace}
\usepackage{caption}
\ifLuaTeX
  \usepackage{selnolig}  
\fi
\usepackage[]{natbib}
\usepackage{xr-hyper}
\usepackage{bookmark}

\IfFileExists{xurl.sty}{\usepackage{xurl}}{} 
\newcommand{\anon}{1}

\graphicspath{{plots/}}
\DeclareMathOperator*{\argmax}{arg\,max}
\DeclareMathOperator*{\argmin}{arg\,min}

\def\be{\begin{equation}}
\def\ee{\end{equation}} 
\def\ben{\begin{equation*}}
\def\een{\end{equation*}}
\def\bea{\begin{eqnarray}}
\def\eea{\end{eqnarray}}
\def\bda{\begin{eqnarray*}}
\def\eda{\end{eqnarray*}}
\numberwithin{equation}{section}

\newtheorem{definition}{Definition}
\newtheorem{theorem}{Theorem}
\newtheorem{proposition}{Proposition}
\newtheorem{corollary}{Corollary}
\newtheorem{assumption}{Assumption}

\newtheorem{example}{Example}
\newtheorem{lemma}{Lemma}
\newtheorem{remark}{Remark}

\newcommand{\norm}[1]{\left\lVert#1\right\rVert}

\theoremstyle{definition}

\def\spacingset#1{\renewcommand{\baselinestretch}{#1}\small\normalsize}

\if0\anon
  \hypersetup{pdfauthor={}}
\fi

\begin{document}
\spacingset{1}

\if1\anon
{
  \title{\bf Metric Subtraction for Non-Euclidean Random Objects with Applications to Residual Analysis}
\author{%
  Jiazhen Xu\orcidlink{0009-0006-7870-0340}$^{1,*}$,
  Takuo Matsubara\orcidlink{0000-0003-2177-3056}$^{2}$,
  Han Lin Shang\orcidlink{0000-0003-1769-6430}$^{1}$\\[0.5em]
    {\small $^{1}$Department of Actuarial Studies and Business Analytics,}\\
  {\small Macquarie University, Australia} \\
  {\small $^{2}$Business School, University of Sydney}
  \date{}
}
  \maketitle
} \fi

\if0\anon
{
  \title{\bf Metric Subtraction for Non-Euclidean Random Objects with Applications to Residual Analysis}
  \date{}
  \maketitle
} \fi

\if1\anon
{
\begingroup
\renewcommand{\thefootnote}{\fnsymbol{footnote}}
\footnotetext[1]{Corresponding author: Department of Actuarial Studies and Business Analytics, Level 7, 4 Eastern Road, Macquarie University, Sydney, NSW 2109, Australia; Email: jiazhen.xu@mq.edu.au.}
\endgroup
} \fi

\if0\anon
{
} \fi

\bigskip
\begin{abstract}
Residual analysis for non-Euclidean regression models is challenging due to the lack of a natural subtraction operation and a corresponding notion of conditionally centered residuals. We introduce a metric subtraction framework that yields residuals in the original response space, with their conditional Fr\'echet mean equal to a fixed reference object when the model is correctly specified. We establish sufficient conditions for this conditional centering property, together with a separate condition ensuring distance preservation, providing a construction principle for suitable metric subtraction operators. The proposed metric subtraction framework accommodates several important classes of metric spaces, including spheres, symmetric positive definite matrices, simplices, and Wasserstein spaces. Building on this framework, we develop a randomized test for detecting remaining conditional location effects and establish its asymptotic size and power against fixed alternatives. We further introduce two complementary diagnostic tools, signed distance plots and distance-profile-based boxplots, to visualize systematic residual patterns and identify atypical observations. For uncertainty quantification, we construct residual-based prediction sets and establish their asymptotic validity. Simulation studies and applications to New York City bicycle usage distributions and U.S. electricity generation compositions demonstrate the practical utility of the proposed framework for model assessment, residual diagnostics, and prediction in non-Euclidean regression. \\

\end{abstract}

\vfill

\newpage
\spacingset{1.8} 

\section{Introduction}

Assessing the adequacy of a regression model requires characterizing the variation that remains after accounting for the conditional location of the response. In Euclidean regression, subtracting the true conditional mean yields residuals with zero conditional mean, providing a natural foundation for diagnosing remaining location effects, heteroscedasticity, and atypical observations. For non-Euclidean responses, such as probability distributions, compositions, and networks, commonly referred to as random objects \citep{marron2021object}, recent methodological advances have enabled regression modeling that considers the geometry of the response space through conditional Fr\'echet means. Notable examples include local and global Fr\'echet regression \citep{petersen2019frechet,chen2022uniform}, varying coefficient and single-index Fr\'echet regression \citep{bhattacharjee2022concurrent, bhattacharjee2023single, ghosal2023frechet}, Fr\'echet random forests \citep{capitaine2024frechet,qiu2024random}, and deep learning approaches for random objects \citep{iao2025deep,zhou2026end}. Despite these advances in modeling conditional location, the absence of a natural subtraction operation in general metric spaces poses a fundamental challenge to constructing residuals that retain the geometric structure of the responses. Developing such a residual representation would allow the unexplained variation to be analyzed using intrinsic statistical tools designed for random objects, providing a basis for model diagnostics and inference beyond conditional location estimation.

Although early work on residuals for directional data established an important precedent for extending residual analysis beyond Euclidean settings \citep{jupp1988residuals}, residual analysis for general non-Euclidean random objects remains relatively underdeveloped. Several possible approaches can be considered, such as representing residuals in an ambient Euclidean space, as in \cite{jupp1988residuals}, \cite{cornea2017regression} and \cite{paine2020spherical} for spherical data and symmetric positive definite matrices, or summarizing them through distances between observed and fitted objects, as in \cite{petersen2019frechet}. However, such representations may discard geometric information and limit the scope of subsequent residual diagnostics and inference. These limitations motivate the construction of object-valued residuals that remain in the original response space, thereby allowing to study residuals through their underlying geometry. Recent work such as \cite{xu2026spherically} and \cite{zhu2026geodesic} provides a geometric interpretation of Euclidean subtraction that offers a basis for such a construction. Specifically, the operation $a-b+0$ can be viewed as transporting the geodesic from $b$ to $a$ to a reference point $0$. This interpretation motivates a notion of metric subtraction based on geodesic transport. A classical example of geodesic transport is based on parallel transport and transvection on smooth manifolds \citep{docarmo1992riemannian,lee2018introduction}, with related constructions recently developed for other metric spaces \citep{kurisu2025regression,song2026adopt,zhu2026geodesic}.

Although metric subtraction provides a geometric basis for constructing object-valued residuals, a fundamental statistical question still remains: under what conditions does such an operation yield a meaningful residual representation, and which features of the unexplained variation does it preserve? We address this question by introducing the conditional centering property. Specifically, metric subtraction transports the geodesic between the conditional Fr\'echet mean and the response to a fixed reference object in the response space. The conditional centering property requires that the resulting residuals have this reference object as their conditional Fr\'echet mean for every predictor value. This requirement generalizes the zero conditional mean property of Euclidean regression and provides a statistical foundation for object-valued residual analysis by ensuring that the removal of the true conditional location eliminates systematic conditional location effects.

We establish sufficient conditions for the conditional centering property, providing a principled basis for constructing statistically meaningful metric subtraction operators. We further derive a sufficient condition for distance preservation, which ensures that the distance between each residual and the reference object aligns with that between the corresponding response and its conditional Fr\'echet mean. While conditional centering provides a foundation for detecting systematic departures from the fitted conditional location, distance preservation enables the assessment of residual dispersion and heteroscedasticity. We illustrate the applicability of these conditions by constructing metric subtraction operators that satisfy both properties in several important object spaces, including spheres, spaces of symmetric positive definite matrices, and simplices.

Having established a framework for constructing conditionally centered object-valued residuals, we turn to their statistical analysis. We develop a suite of inferential, diagnostic, and uncertainty quantification tools, each of which is also of independent interest and can be applied directly to general random objects. We develop a test of the conditional centering property which assesses whether the conditional Fr\'echet mean of the residuals coincides with the reference object across predictor values, motivated by the cost-function comparison approach of \cite{song2026inference}. The test directly targets the true conditional Fr\'echet mean, rather than the global Fr\'echet regression mean considered in \cite{song2026inference}. This distinction is important because the conditional Fr\'echet mean need not be the minimizer of a global Fr\'echet regression function. We establish the asymptotic size and power of the proposed test against fixed alternatives under general high-level conditions, satisfied by concrete implementations based on Nadaraya--Watson Fr\'echet regression for continuous predictors and category-wise Fr\'echet means for categorical predictors.

For residual diagnostics and uncertainty quantification, we introduce three complementary tools. First, we propose a signed distance plot that incorporates directional contrasts, defined through reference landmarks, into conventional distance-based visualizations, thereby revealing systematic location patterns that may be obscured by unsigned distances. Second, building on the distance profiles of \cite{dubey2024metric}, we generalize the Wasserstein boxplot of \cite{liu2026wasserstein} to general metric spaces; we use the distance profile quantiles and resulting center-outward ordering to construct a functional-boxplot-style visualization \citep{HS10, sun2011functional, DG18} for characterizing residual variation and identifying atypical observations. Finally, we combine the quantiles of the mean-distance scores with the metric subtraction operator to construct residual-based prediction sets and establish their asymptotic validity.

Together, these developments provide, to our knowledge, the first general framework for residual analysis of non-Euclidean random objects. The main contributions are summarized as follows:
\begin{itemize}
\item We establish sufficient conditions for the conditional centering property and use them as a construction principle for metric subtraction. Subtracting the true conditional Fr\'echet mean through the resulting operators yields residuals in the response space whose conditional Fr\'echet mean equals a fixed reference object.
\item We propose a test of whether the conditional Fr\'echet mean of the random objects equals the reference object, and establish asymptotic size and consistency against fixed alternatives.
\item Two visualization tools are developed, covering the signed distance plot and a distance-profile-based boxplot for random objects.
\item We construct asymptotically valid residual-based prediction sets for future random objects.
\end{itemize}

The remainder of the paper is organized as follows. Section~\ref{sect::metric subtraction} introduces metric subtraction and its sufficient conditions. Section~\ref{sect::residual analysis} develops the diagnostic, testing, boxplot, and prediction procedures. Sections~\ref{sect::simulation} and \ref{sect::applications} report simulations and data analyses, respectively. Section~\ref{sect::discussion} discusses the scope of the framework and directions for further development. Proofs, supporting properties of the conditional mean estimator, test implementation details, and additional numerical results are provided in the supplementary material.

\section{Metric subtraction operator}\label{sect::metric subtraction}

\begin{figure}[!htb]
\centering
\includegraphics[width=\linewidth]{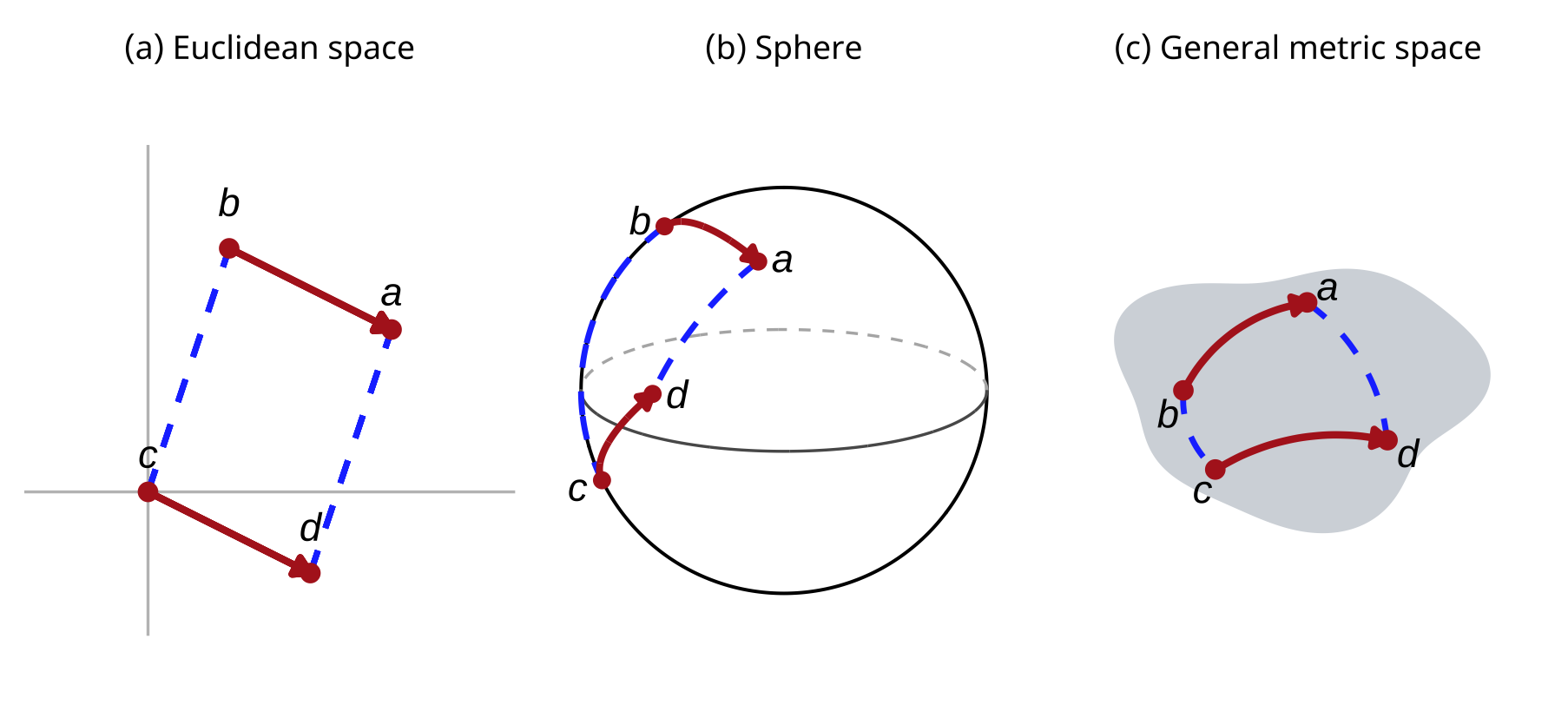}
\caption{Illustration of the geodesic transport designed to subtract $b$ from $a$, given a reference point $c$. (a) in Euclidean space, (b) on the sphere, and (c) in a general metric space. Solid red lines denote the geodesics of $b \to a$ and $c \to d$ with $d=T_{b,a}(c)$, while dashed blue lines indicate the transport of the geodesics between points.}\label{fig::optimal transport visual}
\end{figure}

Let $(\Omega,d)$ be the response space. As noted by \cite{zhu2026geodesic} and \cite{xu2026spherically}, the subtraction $a-b=a-b+0$ in Euclidean space can be interpreted as transporting the geodesic from $b$ to $a$ to a reference point 0; see Figure~\ref{fig::optimal transport visual}(a) for an illustration. Motivated by this idea, for $a,b,c\in\Omega$ we write $T_{b,a}(c)$ for the object obtained by transporting the geodesic from $b$ to $a$ to the reference point $c$; see Figure~\ref{fig::optimal transport visual}(b) and (c) for the corresponding idea on the sphere and in a general metric space. Here, we refer to $T$ as the metric subtraction operator.

Let $(Y,X)$ take values in $\Omega\times\mathcal X$. Suppose that, for every $x\in\mathcal X$ and $\nu\in\Omega$, $\mathbb E\{d^2(\nu,Y)\mid X=x\}<\infty$, and that the conditional Fr\'echet mean $r(x)=\argmin_{\nu\in\Omega}\mathbb E\{d^2(\nu,Y)\mid X=x\}$ exists and is unique. Given a fixed reference point $\omega\in\Omega$, define the object-valued residual $\mathcal E=T_{r(X),Y}(\omega)$. Note that in Euclidean space, the default choice $\omega=0$ yields $\mathcal E=Y-r(X)$ and $\mathbb E(\mathcal E\mid X=x)=0$. For general metric spaces there is no default choice for $\omega$, and the corresponding requirement for object-valued residuals is a constant conditional Fr\'echet mean at $\omega$, given in the following definition.

\begin{definition}\label{def::metric subtraction}
A metric subtraction operator $T:\Omega\times\Omega\times\Omega\to\Omega$ is \emph{suitable for residual construction} for $(Y,X)$ and reference point $\omega$ if, for every $x$, the conditional Fr\'echet mean of $\mathcal E=T_{r(X),Y}(\omega)$ is equal to $\omega$, that is,
\begin{align}\label{formula::conditional centering}
    \omega=\argmin_{\nu\in\Omega}\mathbb E\{d^2(\nu,\mathcal E)\mid X=x\}.
\end{align}
If, in addition,
\begin{align}\label{formula::distance preserve}
d(r(X),Y)=d(\omega,\mathcal E)\qquad\text{almost surely},
\end{align}
then $T$ is called a distance-preserving metric subtraction operator.
\end{definition}

The conditional centering property \eqref{formula::conditional centering} in Definition~\ref{def::metric subtraction} is the object-valued analogue of the zero conditional mean property $\mathbb E(\mathcal E\mid X=x)=0$ in Euclidean regression. This requirement is stronger than the common marginal Fr\'echet mean property, as it ensures that metric subtraction of $r(X)$ removes systematic conditional location effects. The distance-preserving property \eqref{formula::distance preserve} imposes an additional requirement on the metric subtraction operator, ensuring that the distance of each residual from the reference point equals that of the corresponding response from its conditional Fr\'echet mean. Importantly, conditional centering does not imply distance preservation, and certain geodesic transport operators preserve the conditional Fr\'echet mean while distorting distances from the center, as illustrated in subsequent examples in Section~\ref{subsect::examples}.

In the following, we establish sufficient conditions under which the properties in Definition~\ref{def::metric subtraction} hold. We give two sets of conditions. The first requires the residual map $y\mapsto T_{r(x),y}(\omega)$ to be an isometry, as stated in the following assumption.

\begin{assumption}\label{assump:isometry}
For every $x\in\mathcal X$, there is an isometry $g_x$ of $\Omega$ onto itself such that $g_x\{r(x)\}=\omega$ and $T_{r(x),Y}(\omega)=g_x(Y)$ almost surely given $X=x$.
\end{assumption}

The isometry requirement in Assumption~\ref{assump:isometry} holds for several examples discussed later in Section~\ref{subsect::examples}, including spaces that are flat in a global chart, Riemannian symmetric spaces, and the simplex. The second set of conditions covers residual maps that are not isometries. For each $\nu\in\Omega$, let $(\mathcal V_\nu,\langle\cdot,\cdot\rangle_\nu)$ be a separable Hilbert space with norm $\norm{\cdot}_\nu=\langle\cdot,\cdot\rangle_\nu^{1/2}$. Let $\mathcal R_\nu:\operatorname{dom}(\mathcal R_\nu)\to\mathcal V_\nu$ be a Borel map, defined on a set $\operatorname{dom}(\mathcal R_\nu)\subseteq\Omega$ containing $\nu$, such that $\mathcal R_\nu(\nu)=0$ and $d(\nu,q)=\norm{\mathcal R_\nu(q)}_\nu$ for all $q\in\operatorname{dom}(\mathcal R_\nu)$. For every $\nu\in\Omega$, suppose there is an admissible set $\mathcal D_\nu\subseteq\mathcal V_\nu$ and a map $\mathcal R_\nu^{-1}:\mathcal D_\nu\to\operatorname{dom}(\mathcal R_\nu)$ such that $\mathcal R_\nu\{\mathcal R_\nu^{-1}(v)\}=v$ for all $v\in\mathcal D_\nu$. Integrals of $\mathcal V_\nu$-valued maps are Bochner integrals.

\begin{assumption}\label{assump:mean}
For every $x\in\mathcal X$, $Y\in\operatorname{dom}(\mathcal R_{r(x)})$ almost surely given $X=x$, and $\mathbb E\{\mathcal R_{r(x)}(Y)\mid X=x\}=0$.
\end{assumption}

\begin{assumption}\label{assump:linear transport}
For the reference point $\omega$ and every $x\in\mathcal X$, there is a linear operator $\mathcal T_{r(x)\to\omega}:\operatorname{dom}(\mathcal T_{r(x)\to\omega})\to\mathcal V_\omega$, whose domain is a linear subspace of $\mathcal V_{r(x)}$, that commutes with the conditional expectation of $\mathcal R_{r(x)}(Y)$. That is, $\mathcal R_{r(x)}(Y)\in\operatorname{dom}(\mathcal T_{r(x)\to\omega})$ almost surely given $X=x$, $\mathcal T_{r(x)\to\omega}\{\mathcal R_{r(x)}(Y)\}$ is Bochner integrable given $X=x$, $\mathbb E\{\mathcal R_{r(x)}(Y)\mid X=x\}\in\operatorname{dom}(\mathcal T_{r(x)\to\omega})$, and
\[
\mathbb E\left[\mathcal T_{r(x)\to\omega}\{\mathcal R_{r(x)}(Y)\}\mid X=x\right]=\mathcal T_{r(x)\to\omega}\left[\mathbb E\{\mathcal R_{r(x)}(Y)\mid X=x\}\right].
\]
\end{assumption}

Assumption~\ref{assump:mean} is the first-order condition for the conditional Fr\'echet mean, and one sufficient condition for Assumption~\ref{assump:linear transport} is that $\mathcal T_{r(x)\to\omega}$ is bounded. The next theorem shows, under either set of conditions, how to construct a suitable metric subtraction operator $T$ that satisfies the conditional centering property \eqref{formula::conditional centering}.

\begin{theorem}\label{thm:main 2}
Let $\omega\in\Omega$ be fixed and let $\mathcal E=T_{r(X),Y}(\omega)$. Suppose that one of the following holds:
\begin{enumerate}
\item[(i)] Assumption~\ref{assump:isometry}; or
\item[(ii)] Assumptions~\ref{assump:mean} and~\ref{assump:linear transport}, with
\begin{align}\label{formula::metric subtraction}
T_{\nu_1,\nu_2}(\nu_3)=\mathcal R_{\nu_3}^{-1}\left[\mathcal T_{\nu_1\to\nu_3}\{\mathcal R_{\nu_1}(\nu_2)\}\right],
\end{align}
and, for every $x\in\mathcal X$,
\begin{enumerate}
\item[(a)] $\mathcal T_{r(x)\to\omega}\{\mathcal R_{r(x)}(Y)\}\in\mathcal D_\omega$ almost surely given $X=x$; and
\item[(b)] The conditional distribution of $\mathcal E$ given $X=x$ has a finite second moment, and if $\mathbb E\{\mathcal R_\omega(\mathcal E)\mid X=x\}=0$, then $\omega$ is the unique conditional Fr\'echet mean of $\mathcal E$ given $X=x$.
\end{enumerate}
\end{enumerate}
Then, for every $x\in\mathcal X$, $\omega$ is the unique conditional Fr\'echet mean of $\mathcal E$ given $X=x$. In particular, the conditional centering property \eqref{formula::conditional centering} in Definition~\ref{def::metric subtraction} holds.
\end{theorem}

Theorem~\ref{thm:main 2} establishes the conditional centering property based on either set of conditions, using the isometry property or linearity of the transport $\mathcal T_{\nu_1\to\nu_3}$. An additional requirement on norm preservation is needed only if one also wishes to achieve distance preservation \eqref{formula::distance preserve} in Definition~\ref{def::metric subtraction}, as given in the following corollary.

\begin{corollary}\label{cor::distance preservation}
Under Assumption~\ref{assump:isometry}, $d(r(X),Y)=d(\omega,\mathcal E)$ almost surely. On the other hand, suppose that $T$ is given by \eqref{formula::metric subtraction} and that condition~(a) of Theorem~\ref{thm:main 2} holds. If, for every $x$, $\norm{\mathcal T_{r(x)\to\omega}(v)}_\omega=\norm{v}_{r(x)}$ for all $v$ in the support of $\mathcal R_{r(x)}(Y)$, then $d(r(X),Y)=d(\omega,\mathcal E)$ almost surely.
\end{corollary}

\subsection{Examples of metric subtraction}\label{subsect::examples}

Theorem~\ref{thm:main 2} and Corollary~\ref{cor::distance preservation} provide guidelines for constructing the metric subtraction operator. In the following, we provide several examples describing how metric subtraction can be implemented.

\begin{example}[Euclidean space]\label{exmp::1}
Let $\Omega=\mathbb R^d$ with the Euclidean metric $d(\nu_1,\nu_2)=\norm{\nu_1-\nu_2}_{\rm E}$. Construct $T_{\nu_1,\nu_2}(\nu_3)=\nu_2-\nu_1+\nu_3$. 
\end{example}

\begin{example}[Sphere]\label{exmp::2}
Let $\Omega=\mathcal S^{d-1}$, $d\geq 2$, be the unit sphere equipped with the intrinsic metric $d(\nu_1,\nu_2)=\arccos(\nu_1^\top\nu_2)$. For $\nu_3\neq\pm\nu_1$, define the rotation matrix $R_{\nu_1,\nu_3}=I+\sin(\theta_{\nu_1,\nu_3})Q_{\nu_1,\nu_3}+\{1-\cos(\theta_{\nu_1,\nu_3})\}Q_{\nu_1,\nu_3}^2$ \citep{zhu2026geodesic}, where $I$ is the identity, $\theta_{\nu_1,\nu_3}=d(\nu_1,\nu_3)$, $Q_{\nu_1,\nu_3}=u_2u_1^\top-u_1u_2^\top$, $u_1=\nu_1$ and $u_2=\{\nu_3-(\nu_3^\top u_1)u_1\}/\norm{\nu_3-(\nu_3^\top u_1)u_1}_{\rm E}$, and set $R_{\nu_1,\nu_1}=I$. Construct $T_{\nu_1,\nu_2}(\nu_3)=R_{\nu_1,\nu_3}\nu_2$.
\end{example}

\begin{example}[Space of symmetric positive definite matrices]\label{exmp::3}
Let $\Omega=\mathcal S_d^+$ be the space of $d\times d$ symmetric positive definite matrices. Any $S\in\mathcal S_d^+$ is mapped via Cholesky decomposition to a unique matrix $L\in\mathcal L_+$, where $\mathcal L_+$ is the space of lower-triangular matrices with positive diagonals. Let $\lfloor L\rfloor$ denote the strictly lower-triangular part of $L$, and $\mathbb Z(L)$ its diagonal part. Equip $\mathcal L_+$ with the log-Cholesky metric \citep{lin2019riemannian} $d(L_1,L_2)=\left\{\norm{\lfloor L_1\rfloor-\lfloor L_2\rfloor}_F^2+\norm{\log\mathbb Z(L_1)-\log\mathbb Z(L_2)}_F^2\right\}^{1/2}$, where $\norm{\cdot}_F$ is the Frobenius norm. Construct $T_{L_1,L_2}(L_3)=L_4$, where $\lfloor L_4\rfloor=\lfloor L_3\rfloor+(\lfloor L_2\rfloor-\lfloor L_1\rfloor)$ and $\mathbb Z(L_4)=\exp[\log\mathbb Z(L_3)+\{\log\mathbb Z(L_2)-\log\mathbb Z(L_1)\}]$.
\end{example}

\begin{example}[Riemannian symmetric spaces]\label{exmp::4}
Let $\Omega=\mathcal M$ be a Riemannian symmetric space with geodesic distance $d$ and injectivity radius $\operatorname{inj}(\mathcal M)>0$. Each $p\in\mathcal M$ has a geodesic symmetry $s_p$. For $d(\nu_1,\nu_3)<\operatorname{inj}(\mathcal M)$, let $\gamma$ be the unique minimizing geodesic from $\nu_1$ to $\nu_3$, let $\tau_{\nu_1,\nu_3}=s_{\gamma(1/2)}\circ s_{\nu_1}$ be the transvection along $\gamma$ where $\circ$ is ordinary composition of maps \citep[see][page~231]{o1983semi}. Construct $T_{\nu_1,\nu_2}(\nu_3)=\tau_{\nu_1,\nu_3}(\nu_2)$. 
\end{example}

\begin{example}[Space of graph Laplacians]\label{exmp::5}
Let $\Omega=\mathcal L_d^+$ be the space of fully connected graph Laplacians. Any $L\in\mathcal L_d^+$ is uniquely identified by its upper-triangular edge weights $w\in\mathbb R_+^K$, where $K=d(d-1)/2$. Equip $\mathcal L_d^+$ with the metric $d(L_1,L_2)=\left\{\sum_{i<j}\left(\log w_{ij}^{(1)}-\log w_{ij}^{(2)}\right)^2\right\}^{1/2}$. Construct $T_{L_1,L_2}(L_3)=L_4$, where the edge weights of $L_4$ are $w_{ij}^{(4)}=w_{ij}^{(3)}w_{ij}^{(2)}/w_{ij}^{(1)}$. 
\end{example}

\begin{example}[Simplex]\label{exmp::6}
Let $\Omega=\Delta^{d-1}$ be the space of compositions equipped with the Fisher--Rao metric $d(x,y)=\arccos(\sum_{j=1}^d\sqrt{x_jy_j})$. Construct $T_{\nu_1,\nu_2}(\nu_3)=(R_{\sqrt{\nu_1},\sqrt{\nu_3}}\sqrt{\nu_2})^2$, with the rotation of Example~\ref{exmp::2} and $(\cdot)^2$ the componentwise square. To prevent $T$ from moving points outside the simplex, one needs $R_{\sqrt{r(x)},\sqrt{\omega}}\sqrt Y \in S^{d-1}_{+}$ conditionally almost surely for every~$x$, where $S^{d-1}_{+}$ is the positive orthant of the sphere.
\end{example}

Examples~\ref{exmp::1}--\ref{exmp::6} are the metric spaces where the isometry requirement in Assumption~\ref{assump:isometry} holds. In the following, we give examples where the second set of conditions, Assumptions~\ref{assump:mean} and \ref{assump:linear transport}, holds.  

\begin{example}[Riemannian manifolds]\label{exmp::7}
Let $\Omega=\mathcal M$ be a complete Riemannian manifold with geodesic distance $d$, injectivity radius $\operatorname{inj}(\mathcal M)>0$, and sectional curvatures bounded above by $\vartheta\in\mathbb R$.  Let $\mathcal V_\nu$ be the tangent space at $\nu$, let $\mathcal R_\nu=\operatorname{Log}_\nu$ be the logarithmic map on $\operatorname{dom}(\mathcal R_\nu)=\{q\in\mathcal M:d(\nu,q)<\operatorname{inj}(\mathcal M)\}$, and let $\mathcal R_\nu^{-1}=\operatorname{Exp}_\nu$ be the exponential map on $\mathcal D_\nu=\{v\in\mathcal V_\nu:\norm{v}_\nu<\operatorname{inj}(\mathcal M)\}$. For $d(\nu_1,\nu_3)<\operatorname{inj}(\mathcal M)$, let $\mathcal T_{\nu_1\to\nu_3}=\Gamma_{\nu_1\to\nu_3}$ be the parallel transport along the unique minimizing geodesic from $\nu_1$ to $\nu_3$. Then \eqref{formula::metric subtraction} gives $T_{\nu_1,\nu_2}(\nu_3)=\operatorname{Exp}_{\nu_3}\left[\Gamma_{\nu_1\to\nu_3}\{\operatorname{Log}_{\nu_1}(\nu_2)\}\right]$. 
\end{example}

\begin{example}[Wasserstein space of univariate distributions]\label{exmp::8}
Let $\Omega=\mathcal W_2$ be a class of univariate absolutely continuous probability measures supported on a common bounded interval, equipped with the 2-Wasserstein metric $d(\nu_1,\nu_2)=\norm{Q_{\nu_1}-Q_{\nu_2}}_{L^2}=\left[\int_0^1\{Q_{\nu_1}(u)-Q_{\nu_2}(u)\}^2\,du\right]^{1/2}$, where $F_{\nu}$ and $Q_{\nu}=F_{\nu}^{-1}$ denote the cumulative distribution function (CDF) and quantile function of $\nu\in \mathcal W_2$, respectively. Let $\mathcal V_\nu=L^2(0,1)$ and $\mathcal R_\nu(y)=Q_y-Q_\nu$, with $\mathcal R_\nu^{-1}(v)$ mapping a quantile function $Q_\nu+v$ back to its probability measure. For $\nu_1,\nu_3\in\Omega$, define $\phi_{\nu_1,\nu_3}(u)=F_{\nu_1}\{Q_{\nu_3}(u)\}$ and $\mathcal T_{\nu_1\to\nu_3}(v)=v\circ\phi_{\nu_1,\nu_3}$. Assume the CDFs are continuous and strictly increasing on the common support. Then (\ref{formula::metric subtraction}) gives $T_{\nu_1,\nu_2}(\nu_3)=(Q_{\nu_2}\circ F_{\nu_1})_\#\nu_3$, where $(Q_{\nu_2}\circ F_{\nu_1})_\#\nu_3$ denotes the pushforward of the distribution $\nu_3$ by the map $Q_{\nu_2}\circ F_{\nu_1}$.
\end{example}

The verification of the sufficient conditions in Theorem~\ref{thm:main 2} and/or Corollary~\ref{cor::distance preservation} for Examples~\ref{exmp::1}--\ref{exmp::8} is provided in Section~S1 of the supplementary material.

\section{Diagnostic, testing, and uncertainty quantification}\label{sect::residual analysis}

Let $(\Xi, \mathcal{F}, \mathbb{P})$ be a probability space large enough for a stochastic process $\{(Y_i, X_i)\}_{i=1}^\infty \subset \Omega \times \mathcal{X}$ to be measurable. All probabilities in the following are with respect to $(\Xi, \mathcal{F}, \mathbb{P})$ unless stated otherwise. Suppose that we have a sequence of observations $\{(Y_i, X_i)\}_{i=1}^n \subset \Omega \times \mathcal{X}$ and $\mu_Y$ is the population Fr\'{e}chet mean of $\{Y_i\}_{i=1}^n$, assumed to exist and be unique. Throughout this section, we use $\omega=\mu_Y$ as the reference object and define the residuals by $\mathcal E_i=T_{r(X_i),Y_i}(\mu_Y)$, $i=1,2,\ldots,n$. The choice $\omega=\mu_Y$ centers the residuals at an object with a population interpretation. A different fixed reference point can be used when it is more natural for the application. The residuals $\{\mathcal E_i\}_{i=1}^n$ are further assumed to be independent and identically distributed (i.i.d.) copies of the random object $\mathcal E$ and take values in a totally bounded subset of $\Omega$. Note that this assumption is separate from the conditional centering property, and the original observations $\{(Y_i, X_i)\}_{i=1}^n$ can be dependent. Under Theorem~\ref{thm:main 2}, the conditional Fr\'{e}chet mean of $\mathcal E$ given $X=x$ is $\mu_Y$ when the working model $r$ equals the true conditional Fr\'{e}chet mean of $Y$, and it is the analogue of the zero conditional mean property $\mathbb E(\mathcal E\mid X=x)=0$ for Euclidean regression, where $Y=r(X)+\mathcal E$.  Define its conditional dispersion about the reference point by $V_{\mathcal E}(x)=\mathbb E\{d^2(\mu_Y,\mathcal E)\mid X=x\}$. The additional condition $V_{\mathcal E}(x)\equiv V$ is an object-valued analogue of homoscedasticity. 

Section~\ref{sect::sign dist plot} introduces the signed distance plot, Section~\ref{sect::residual test} develops the conditional Fr\'{e}chet mean test, Section~\ref{sect::boxplot} presents the boxplot, and Section~\ref{sect::PI} constructs prediction sets. These procedures can apply to both the original observations $\{(Y_i,X_i)\}_{i=1}^n$ and object-valued residuals, and their application to residuals constitutes residual analysis. To streamline notation, we formulate the methods using the oracle residual sample $\{\mathcal E_i\}_{i=1}^n$, and the theoretical analysis focuses primarily on the oracle residual setting. In practice, the regression function $r(\cdot)$ and the population Fr\'echet mean $\mu_Y$ are unknown. Given a fitted model $\widehat r:\mathcal X\to\Omega$ and the sample Fr\'echet mean $\widehat\mu_Y=\argmin_{\nu\in\Omega}n^{-1}\sum_{i=1}^n d^2(\nu,Y_i)$, the fitted residual is $\widehat{\mathcal E}_i=T_{\widehat r(X_i),Y_i}(\widehat\mu_Y)$. 

\subsection{Signed distance plot}\label{sect::sign dist plot}

In classical linear regression, scatter plots of residuals against predictors provide a useful diagnostic for identifying systematic patterns, where residuals can take both positive and negative values around zero. For non-Euclidean random objects, however, an analogous visualization is not directly available. A natural approach is to plot the distances between residuals and a fixed reference object, but these distances are always nonnegative and therefore cannot distinguish deviations in different directions from the reference. Consequently, systematic location patterns that would be apparent in a conventional residual scatter plot may be obscured. To address this limitation, we introduce a signed distance plot that assigns signs to residual distances, providing a scalar visualization analogous to the classical residual scatter plot while retaining information about the magnitude of deviations. The signed distance plot can be plotted against a univariate predictor, displayed by predictor category for regression settings, or ordered by observation index or time. For a general multivariate or object-valued predictor, an index plot describes patterns along the chosen ordering.

Choose $a_-\in\Omega$ and then select $a_+$ such that $d(a_-,\mu_Y)=d(a_+,\mu_Y)$ and $a_-,a_+$, and $\mu_Y$ lie on the same geodesic. This can be achieved via $a_+=T_{a_-,\mu_Y}(\mu_Y)$ when the chosen metric subtraction operator $T$ is distance-preserving, or using the geodesic extension following equation (2) in \cite{zhu2026geodesic} when $T$ is not distance-preserving. Define $S(y)=d^2(y,a_-)-d^2(y,a_+)$, so that $S(y)>0$ when $y$ is relatively closer to $a_+$. The signed distance is $D_i={\rm sign}\{S(\mathcal E_i)\}d(\mathcal E_i,\mu_Y)$ where ${\rm sign}(\cdot)$ is the sign function. This construction contains two pieces of information, where the magnitude measures departure from the residual center, while the sign records position relative to the chosen points $a_-$ and $a_+$. The following theorem identifies conditions under which the signed distances have zero mean and constant variance.

\begin{theorem}\label{thm::residual distance plot}
For $i=1,2,\ldots,n$, let $\mu_{R,i}$ and $V_i=\mathbb E\{d^2(\mathcal E_i,\mu_{R,i})\}$ denote the Fr\'echet mean and variance of $\mathcal E_i$, respectively. Suppose that $\mu_{R,i}\equiv\mu_Y$ and $V_i\equiv V$ for some constant $V$, and assume $\mathbb P\{S(\mathcal E_i)=0\}=0$. If
\begin{align}\label{formula::symmetry}
\mathbb E\left[d(\mathcal E,\mu_Y)\mathbb I\{S(\mathcal E)>0\}\right]
=
\mathbb E\left[d(\mathcal E,\mu_Y)\mathbb I\{S(\mathcal E)<0\}\right],
\end{align}
where $\mathbb I(\cdot)$ is the indicator function, then $\mathbb E(D_i)\equiv 0$ and ${\rm Var}(D_i)\equiv V$.
\end{theorem}

Condition~\eqref{formula::symmetry} is a directional balance condition around the residual center $\mu_Y$. If the distribution of $\mathcal E$ is symmetric in the sense that $\mathbb P\{S(\mathcal E)>0\}=\mathbb P\{S(\mathcal E)<0\}=1/2$, then it reduces to equality of the mean distances from $\mu_Y$ on the two sides. Therefore, constant location and dispersion of $\{\mathcal E_i\}_{i=1}^n$, together with directional balance, imply a signed distance with mean zero and constant variance.

Note that the signed distance $D_i$ represents an object-valued residual as a scalar and consequently could lose certain information. The distance plot is therefore intended as a first-step visualization, and visible patterns can reveal substantial location effects, periodicity, changes, and atypical observations. Moreover, when the plot is applied to fitted residuals, misspecification of the fitted model $\widehat r$ may induce visible structure. In practice, one may take $a_-=\argmax_{\mathcal E_i\in\{\mathcal E_1,\mathcal E_2,\ldots,\mathcal E_n\}}d^2(\mathcal E_i,\widehat{\mu}_Y)$ when no extreme outlying object exists, or choose $a_-$ as the residual based on mean-distance scores defined in Section~\ref{sect::boxplot} for robustness.

\begin{figure}[!htb]
\centering
\includegraphics[width=\linewidth]{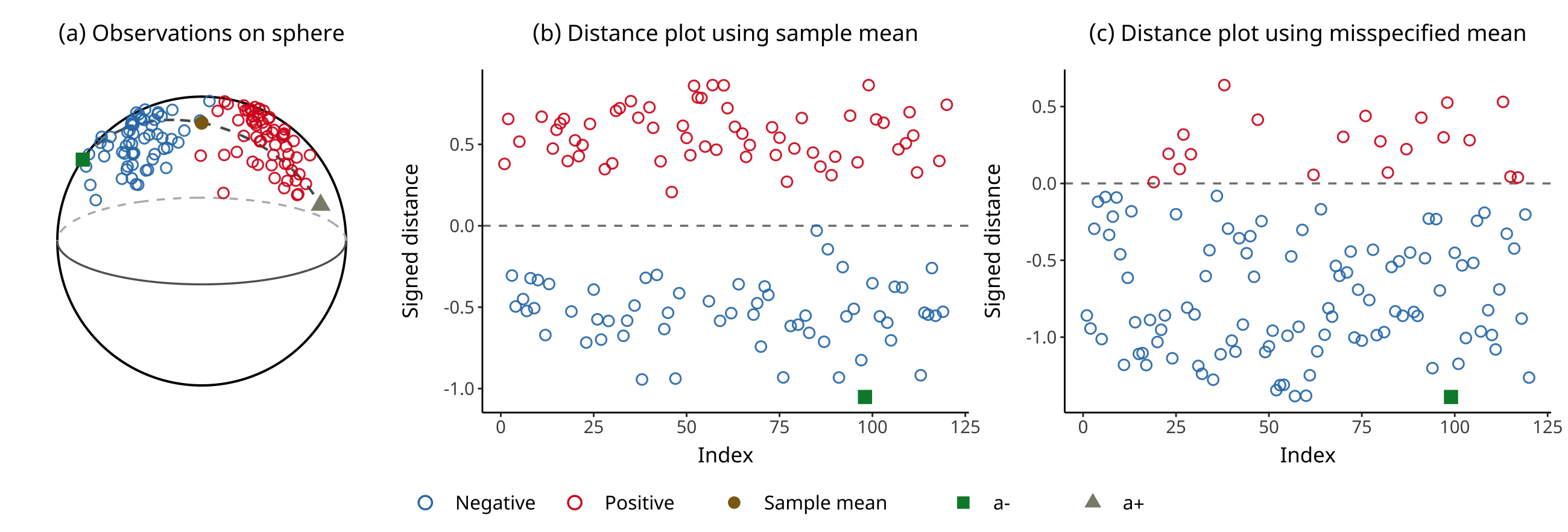}
\caption{(a) Illustration of simulated spherical observations, (b) signed distance plot using the sample mean, (c) signed distance plot using the misspecified mean. Blue and red circles correspond to observations with negative and positive distances, respectively.}\label{fig::residual distanct plot}
\end{figure}

To illustrate the proposed signed distance, we consider two settings of spherical data generation, and present distance plots for the simulated data. Figure~\ref{fig::residual distanct plot}(a) shows spherical observations from two groups with different centers and the same dispersion. The observations are colored according to the sign of the signed distance using the sample mean of the simulated data. Figure~\ref{fig::residual distanct plot}(b) presents the signed distances, and the dispersion of the signed residuals appears constant. Figure~\ref{fig::residual distanct plot}(c) uses a misspecified mean, and it shows that replacing the sample mean by a misspecified mean leads to signed distances whose mean is no longer centered at zero.

\begin{figure}[!htb]
\centering
\includegraphics[width=\linewidth]{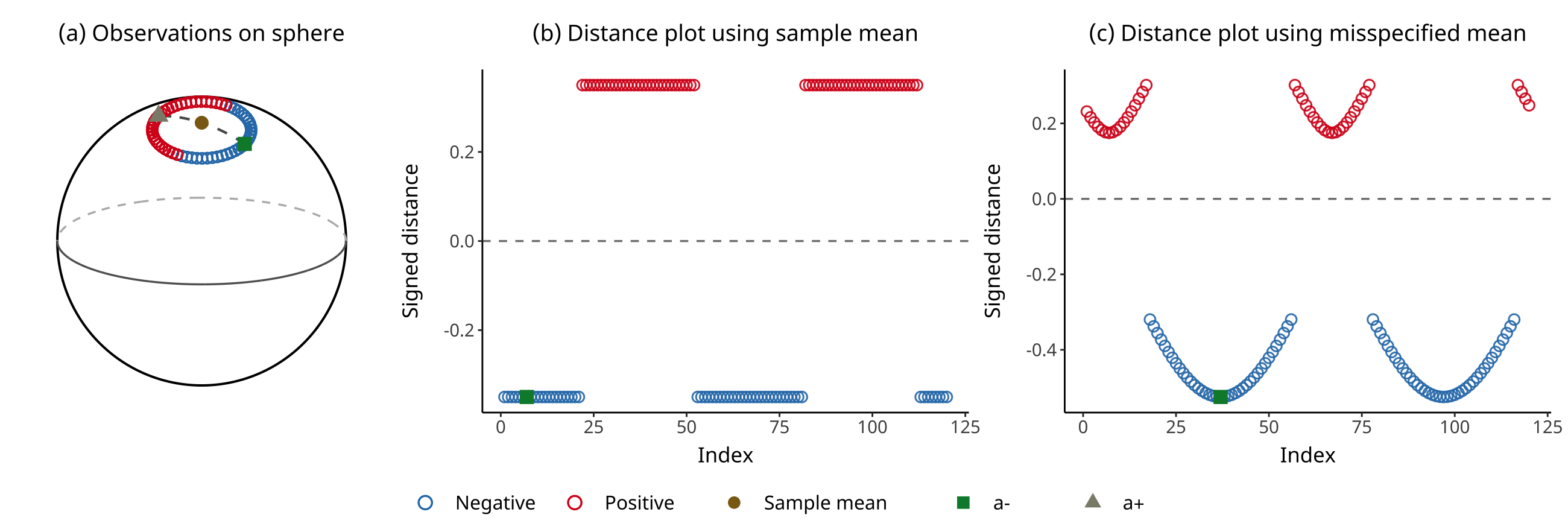}
\caption{(a) Illustration of nonrandom periodic spherical observations, (b) signed distance plot using the sample mean, (c) signed distance plot using the misspecified mean. Blue and red circles correspond to observations with negative and positive distances, respectively.}\label{fig::residual distanct plot periodic}
\end{figure}

We next illustrate that the distance plot can also display periodicity. Figure~\ref{fig::residual distanct plot periodic}(a) presents simulated deterministic spherical observations with period 60 and a constant distance from the center. Figure~\ref{fig::residual distanct plot periodic}(b) exhibits a clear periodic pattern in the signed distances even though the unsigned distance is constant. When a misspecified mean is used in Figure~\ref{fig::residual distanct plot periodic}(c), the resulting signed distances fail to reflect the underlying constant distance to the center in the data-generating process.

\subsection{Conditional Fr\'echet mean test}\label{sect::residual test}

Although the signed distance plot introduced in Section~\ref{sect::sign dist plot} provides a graphical diagnostic for residual patterns, a formal inferential procedure is needed to assess whether the model $r$ adequately captures the systematic conditional location effects. Denote the conditional Fr\'echet mean of the residual $\mathcal E$ by $m(x):=\argmin_{\nu\in\Omega} M(\nu,x)$ where $M(\nu,x)=\mathbb E\{d^2(\nu,\mathcal E)\mid X=x\}$. Suppose that $m(x)$ exists and is unique. Then, following Theorem~\ref{thm:main 2}, we aim to test the null hypothesis
\begin{align}\label{formula::null hypothesis}
    {\rm H}_0: \mathbb P_X( m(X)=\mu_Y) =1,
\end{align}
versus the alternative hypothesis
\begin{align}\label{formula::alter hypothesis}
    {\rm H}_1: \mathbb P_X( m(X)\neq \mu_Y) >0.
\end{align}
When the null hypothesis is rejected, it indicates that the population target of the working model used to construct the residuals may differ from $r$, and hence that the model is underfitted.

Suppose we have a consistent estimator of $m(x)$ as $\widehat m(x) = \argmin_{\nu \in \Omega} \widehat M(\nu,x)$ with some cost function $\widehat M$. Set $A_{ni}=\widehat M(\mu_Y,X_i)$, $B_{ni}=\widehat M\{\widehat m(X_i),X_i\}$, and $\widehat D_n=n^{-1}\sum_{i=1}^n(A_{ni}-B_{ni})$. The difference $A_{ni}-B_{ni}$ is nonnegative and measures the improvement in the estimated objective from replacing the constant reference $\omega=\mu_Y$ by a predictor-specific location. Its average is negligible under the null when estimation error is sufficiently controlled, and has a positive population limit under a fixed alternative.

Motivated by objective comparisons in global Fr\'echet regression inference \citep{song2026inference}, we consider to adopt the independent multiplier method \citep{gine1990bootstrapping}. Let $\phi_1,\ldots,\phi_n$ and $\psi_1,\ldots,\psi_n$ be mutually independent and independent of the data, with $\mathbb E(\phi_i)=\mathbb E(\psi_i)=1$, $\operatorname{Var}(\phi_i)=\operatorname{Var}(\psi_i)=1/2$, and finite $(2+\eta)$th moments for some $\eta>0$. The test statistic is
\begin{align}\label{formula::test stat}
    \widehat Z_n = \sqrt n \widehat T_n / \widehat\sigma_n
\end{align}
where $\widehat T_n=n^{-1}\sum_{i=1}^n(\phi_iA_{ni}-\psi_iB_{ni})$ and $\widehat\sigma_n^2=(2n)^{-1}\sum_{i=1}^n(A_{ni}^2+B_{ni}^2)$.

The next theorem shows that the test rejects the null hypothesis \eqref{formula::null hypothesis} at level $\alpha$ when $\widehat Z_n>z_{1-\alpha}$ where $z_{1-\alpha}$ is the $(1-\alpha)$-quantile of the standard normal distribution.

\begin{theorem}\label{thm::size}
Under the null hypothesis in \eqref{formula::null hypothesis}, suppose that 
\begin{align}\label{formula::high level condition 1}
    \sqrt{n} \widehat D_n = o_p(1), \text{ and }\; \widehat\sigma_n^2  \xrightarrow{p} \sigma^2
\end{align}
where $\sigma^2=\mathbb E [\{M(\mu_Y,X) \}^2]>0$. Then $\widehat Z_n\xrightarrow{d}N(0,1)$.
\end{theorem}

Theorem~\ref{thm::size}  yields $\mathbb P_{{\rm H}_0}(\widehat Z_n>z_{1-\alpha})\longrightarrow\alpha$. The first condition in \eqref{formula::high level condition 1} requires the empirical objective improvement under the null to be negligible at the $n^{-1/2}$ scale, while the second variance condition in \eqref{formula::high level condition 1} ensures a positive limiting scale for the multiplier component. 

We next study the power of the proposed test in \eqref{formula::test stat} given in the following theorem. 

\begin{theorem}\label{thm::power}
Under the alternative hypothesis in \eqref{formula::alter hypothesis}, suppose that 
\begin{align}\label{formula::high level condition 2}
    \widehat D_n - \mathbb E \left\{ M(\mu_Y,X) - M(m(X),X) \right\} = o_p(1), \text{ and } \; \widehat\sigma_n^2 =O_p(1).
\end{align}
Then $\mathbb P_{{\rm H}_1}(\widehat Z_n>z_{1-\alpha})\longrightarrow 1$.
\end{theorem}

Under the alternative, the expected objective improvement is positive on average. The condition~\eqref{formula::high level condition 2} ensures that $\widehat D_n$ estimates that improvement consistently, leading to rejection probability tending to one. Applying these theoretical results to fitted residuals requires control of the discrepancy between objectives computed from fitted and true residuals, as detailed in the following remark.

\begin{remark}\label{remark::fitted residuals}
Let $\widehat{M}_{n}^{f}$ denote the objective constructed from the fitted residuals $\{\widehat{\mathcal{E}}_i\}_{i=1}^{n}$, with its minimizer recomputed accordingly, and set $A_{ni}^f=\widehat M_n^f(\widehat\mu_Y,X_i)$, $B_{ni}^f=\inf_{\nu\in\Omega}\widehat M_n^f(\nu,X_i)$, $\widehat D_n^f=n^{-1}\sum_{i=1}^n(A_{ni}^f-B_{ni}^f)$, and $\widehat\sigma_n^{2,f}=(2n)^{-1}\sum_{i=1}^n\{(A_{ni}^f)^2+(B_{ni}^f)^2\}$. Sufficient conditions for transferring the conclusions of Theorems~\ref{thm::size} and \ref{thm::power} to the fitted residual setting are $\sqrt n\widehat D_n^f/\widehat\sigma_n^{2,f}=o_p(1)$ under the hypothesis in \eqref{formula::null hypothesis}, and $\sqrt n\,\widehat D_n^{\,f}/\widehat\sigma_n^{\,f}\xrightarrow{p}+\infty$ under the alternative hypothesis in~\eqref{formula::alter hypothesis}.
\end{remark}

The sufficient conditions in Remark~\ref{remark::fitted residuals} can be verified based on the explicit metric subtraction operator and the convergence rate of the fitted model $\widehat r$. Borrowing the arguments in Section~S2 of the supplementary material, one can show that the established convergence rates of various models, such as local and global Fr\'echet regression \citep{petersen2019frechet,chen2022uniform}, are sufficient. Additional analysis of the explicit metric subtraction operator can follow the analysis of Assumption 4 (B) in \cite{zhu2026geodesic}. 

Note that the multiplier test in \eqref{formula::test stat} can be repeated and the resulting $p$-values combined using the Cauchy transformation of \citet{liu2020cauchy}. Fix an integer $K\geq1$. For $k=1,\ldots,K$, following the multiplier generation used in the test \eqref{formula::test stat}, we generate $\{\phi_{ik},\psi_{ik}:i=1,\ldots,n\}$ that are independent over all indices and independent of the data, and obtain $\widehat Z_{n,k} = \sqrt n \widehat T_{n,k} / \widehat\sigma_n$ where $\widehat T_{n,k}=n^{-1}\sum_{i=1}^n(\phi_{ik}A_{ni}-\psi_{ik}B_{ni})$. The Cauchy combination test statistic is then given by
\begin{align}\label{formula::Cauchy test}
\widehat C_n=\sum_{k=1}^K c_k \tan\left[\pi\left\{\Phi(\widehat Z_{n,k})-\frac12\right\}\right],
\end{align}
where $\{c_k\}_{k=1}^K$ are deterministic nonnegative weights with $\sum_{k=1}^K c_k=1$, $\tan(\cdot)$ is the tangent, and $\Phi(\cdot)$ is the standard normal CDF. In practice, one may choose $c_k\equiv 1/K$. The next corollary shows that rejecting~\eqref{formula::null hypothesis} when $\widehat C_n>\cot(\pi\alpha)$, where $\cot(\cdot)$ denotes the cotangent, gives a test of asymptotic level $\alpha$.

\begin{corollary}\label{cor::cauchy test size}
If the conditions of Theorem~\ref{thm::size} hold, then $\mathbb P_{{\rm H}_0}(\widehat C_n>\cot(\pi\alpha))\longrightarrow\alpha$.
\end{corollary}

The power of the proposed test in \eqref{formula::Cauchy test} is given in the next corollary. 

\begin{corollary}\label{cor::cauchy test power}
Suppose the conditions of Theorem~\ref{thm::power} hold. Then $\mathbb P_{{\rm H}_1}(\widehat C_n>\cot(\pi\alpha))\longrightarrow 1$.
\end{corollary}

The validity of the test statistic, either \eqref{formula::test stat} or \eqref{formula::Cauchy test}, is built upon the high-level conditions \eqref{formula::high level condition 1} and \eqref{formula::high level condition 2}. Several models can be considered, such as the local Fr\'echet regression in \cite{petersen2019frechet} and the single index Fr\'echet regression in \cite{bhattacharjee2023single,ghosal2023frechet}. Notably, because the population target of global Fr\'echet regression need not equal the conditional Fr\'echet mean \citep{petersen2019frechet,hong2026uniform}, the hypothesis test of \cite{song2026inference} built upon it cannot be applied here.

\subsubsection{Implementation details}

In the following, we illustrate how to construct the test statistic based on a given cost function~$\widehat M$. Here we focus on the case where the predictor space $\mathcal{X}$ is a compact subset of the Euclidean space. The predictor $X$ is further assumed to be univariate and continuous and the Nadaraya--Watson Fr\'echet regression is employed for the analysis. The extension to the multivariate and continuous setting or the non-Euclidean setting is mathematically straightforward, and further details can be found in \cite{dong2026variable} and \cite{schafgans2026multivariate}. For the discrete or categorical setting, one can simply compute the estimator $\widehat m(x)$ via the Fr\'echet mean for each category of $X$, with details in Section~S3 of the supplementary material.

Let $K_h(\cdot)=K(\cdot/h)/h$ for a smoothing kernel $K(\cdot)$ and a bandwidth sequence $h>0$. Assume $\mathbb E\{K_h(X-x)\}>0$ for every $x$ and define the population counterpart of the Nadaraya--Watson Fr\'echet regression estimator as
\begin{align*}
\widetilde{m}(x)=\arg\min_{\nu\in\Omega}\widetilde M(\nu,x),\quad \widetilde M(\nu,x) =  \mathbb E\{K_h(X-x)d^2(\nu,\mathcal E)\} / \mathbb E\{K_h(X-x)\}.
\end{align*}

The next lemma supports the choice of a nonnegative kernel used in the Nadaraya--Watson Fr\'echet regression as it preserves the null for every fixed bandwidth. For the local Fr\'echet regression \citep{petersen2019frechet} where the weights can be negative, this property does not hold.

\begin{lemma}\label{lemma::null preservation}
Under the null hypothesis in \eqref{formula::null hypothesis}, for every $x\in\mathcal X$ and $\nu\in\Omega$, $\widetilde M(\nu,x)-\widetilde M(\mu_Y,x)\geq0$, and $\widetilde m(x)=\mu_Y$ for every $x$ and every $h>0$.
\end{lemma}

For $D_h=\mathbb E_X\left[\widetilde M(\mu_Y,X)-\widetilde M\{\widetilde m(X),X\}\right]$, Lemma~\ref{lemma::null preservation} implies $D_h=0$ under the null hypothesis in \eqref{formula::null hypothesis} for every bandwidth $h>0$. For the empirical implementation, construct the empirical Nadaraya--Watson Fr\'echet regression estimator given by
\begin{align*}
    \widehat m(x)=\arg\min_{\nu\in\Omega}\widehat M(\nu,x),\quad \widehat M(\nu,x) =  \frac{n^{-1}\sum_{j=1}^nK_h(X_j-x)d^2(\nu,\mathcal E_j)}{n^{-1}\sum_{j=1}^nK_h(X_j-x)}.
\end{align*}
 
The verification of the high-level conditions \eqref{formula::high level condition 1} and \eqref{formula::high level condition 2} for the  Nadaraya--Watson Fr\'echet regression setting, and a discussion on the case when $\mathcal{X}$ is not Euclidean can be found in Section~S2 of the supplementary material.

\subsection{Boxplot and outlier detection}\label{sect::boxplot}

This section develops boxplots for random objects $\{\mathcal E_i\}_{i=1}^n$. A boxplot needs a center, nested central regions and a fence beyond which observations are flagged. However, a general metric space supplies none of them directly, because it carries neither a linear order nor coordinatewise quantiles.
We rebuild analogous quantities from distance profiles \citep{dubey2024metric}, following the construction of the Wasserstein boxplot that \citet{liu2026wasserstein} introduced for samples of densities under the Wasserstein metric. We extend their construction from densities to random objects in a general metric space.

\subsubsection{Distance profile}

A distance profile of an object \citep{dubey2024metric} is a distribution of its distances to the population. For a fixed object $z\in\Omega$, the population distance profile, defined by
\begin{align}\label{eq:population_distance_profile}
G_z(t)=\mathbb P\{d(z,\mathcal E)\le t\},\qquad t\geq 0,
\end{align}
is the CDF of the scalar random variable $d(z,\mathcal E)$. Given the residuals $\{\mathcal E_i\}_{i=1}^n$, the empirical profile of 
the $i$th residual $\mathcal E_{i}$ is defined by $\widehat G_i(t)=(n-1)^{-1}\sum_{j\ne i}\mathbb I\{d(\mathcal E_i,\mathcal E_j)\leq t\}$, $t\geq 0$. In the following, for a generic distribution function $G$ on $[0,\infty)$, its quantile function is defined by $G^{-1}(u)=\inf\{t\geq 0:G(t)\geq u\}$ for $u\in(0,1)$.

A central object has small distances to the population, meaning that its profile $G_z$ puts more mass near zero than the profile $G_{\mathcal E}$ of a randomly drawn object.
If so, its quantile function $G_z^{-1}$ lies below $G_{\mathcal E}^{-1}$.
The difference of the two quantile functions, averaged over the random object $\mathcal E$, is the centrality score of \citet{dubey2024metric}, defined as follows:
\begin{align}\label{eq:population_centrality}
C(z)=\mathbb E\left[\int_0^1\left\{G_{\mathcal E}^{-1}(u)-G_z^{-1}(u)\right\}du\right],\qquad z\in\Omega.
\end{align}
This score is large for a central object and small or negative for an outlying object.
\citet{dubey2024metric} rescale the score to $(0,1)$ through the expit function and call the result $\exp\{C(z)\}/[1+\exp\{C(z)\}]$ the transport rank of $z$.
The transport rank orders the objects from the most central to the most outlying.
The sample version of the score averages the quantile differences over the other residuals as
\begin{align*}
\widehat C_i=\frac{1}{n-1}\sum_{j\ne i}\int_0^1\left\{\widehat G_j^{-1}(u)-\widehat G_i^{-1}(u)\right\}du,\qquad i=1,2,\ldots,n .
\end{align*}
The integral of a quantile function is the mean of the corresponding distribution, so the score reduces to a difference of mean distances. To show this, let $\Delta(z)=\mathbb E\{d(z,\mathcal E)\}$ and $\widehat\Delta_i=(n-1)^{-1}\sum_{j\neq i}d(\mathcal E_i,\mathcal E_j)$ for the mean distance from $z$ to the population and its sample version for the $i$th residual, respectively. The next proposition identifies the score with the mean distance, and implies that the most central object is the Fr\'echet median.

\begin{proposition}\label{prop::boxplot}
Let $\mathcal E'$ be an independent copy of $\mathcal E$.
Then, for every $z\in\Omega$, $C(z)=\mathbb E\{d(\mathcal E,\mathcal E')\}-\Delta(z)$. 
For $i=1,2,\ldots,n$, the empirical score satisfies $\widehat C_i=(n-1)^{-1}\sum_{j\ne i}(\widehat\Delta_j-\widehat\Delta_i)$ and
\begin{align}\label{eq::equivalent orderings}
\widehat C_i>\widehat C_j\quad\Longleftrightarrow\quad\widehat\Delta_i<\widehat\Delta_j .
\end{align}
\end{proposition}

By Proposition~\ref{prop::boxplot}, ranking residuals by $\widehat C_i$ from the largest down is ranking them by $\widehat\Delta_i$ from the smallest up. 
The empirical mean-distance score $\widehat\Delta_i$ is small for a central residual and large for an outlying one. Let $\pi$ be a permutation of $\{1,2,\ldots,n\}$ such that $\widehat\Delta_{\pi(1)}\le\dots\le\widehat\Delta_{\pi(n)}$. We write $\mathcal E_{[i]}=\mathcal E_{\pi(i)}$ and $\widehat\Delta_{[i]}=\widehat\Delta_{\pi(i)}$ for the residuals and their scores. The most central residual $\mathcal E_{[1]}$ is the medoid of the sample, and its score $\widehat\Delta_{[1]}$ is the smallest mean distance among the residuals. For $\alpha\in(0,1)$, let $\lceil\alpha n\rceil$ denote the ceiling of $\alpha n$. We define the empirical $\alpha$-central set by $\widehat{\mathcal C}_\alpha=\left\{\mathcal E_{[1]},\ldots,\mathcal E_{[\lceil\alpha n\rceil]}\right\}$, which collects the $\lceil\alpha n\rceil$ most central residuals. The set $\widehat{\mathcal C}_\alpha$ consists of residuals with $\widehat\Delta_i\le\widehat q_\alpha$, 
where $\widehat q_\alpha=\widehat\Delta_{[\lceil\alpha n\rceil]}$ is the empirical $\alpha$-quantile of the scores. The sets are nested in $\alpha$, and $\widehat{\mathcal C}_{1/4}\subseteq\widehat{\mathcal C}_{1/2}\subseteq\widehat{\mathcal C}_{3/4}$ play the roles of the inner box and of the two wider central regions of a classical boxplot.

Their population counterparts are the sublevel sets of the mean-distance function. Let $F_\Delta(t)=\mathbb P\{\Delta(\mathcal E)\le t\}$ be the distribution function of the population mean distance and $q_\alpha=F_\Delta^{-1}(\alpha)$ be its $\alpha$-quantile. The population $\alpha$-central set is then defined by $\mathcal C_\alpha=\{z\in\Omega:\Delta(z)\le q_\alpha\}$, $\alpha\in(0,1)$, which contains $\mathcal E$ with probability at least $\alpha$, and with probability exactly $\alpha$ when $F_\Delta$ is continuous at $q_\alpha$. The population center is a minimizer of $\Delta(\cdot)$, a Fr\'echet median of $\mathcal E$. Note that it exists when $\Omega$ is compact, because $\Delta(\cdot)$ is $1$-Lipschitz, and the sample center $\mathcal E_{[1]}$ always exists.

\subsubsection{Boxplot}\label{sec:3.3.2}

\begin{figure}[!htb]
\centering
\includegraphics[width=\linewidth]{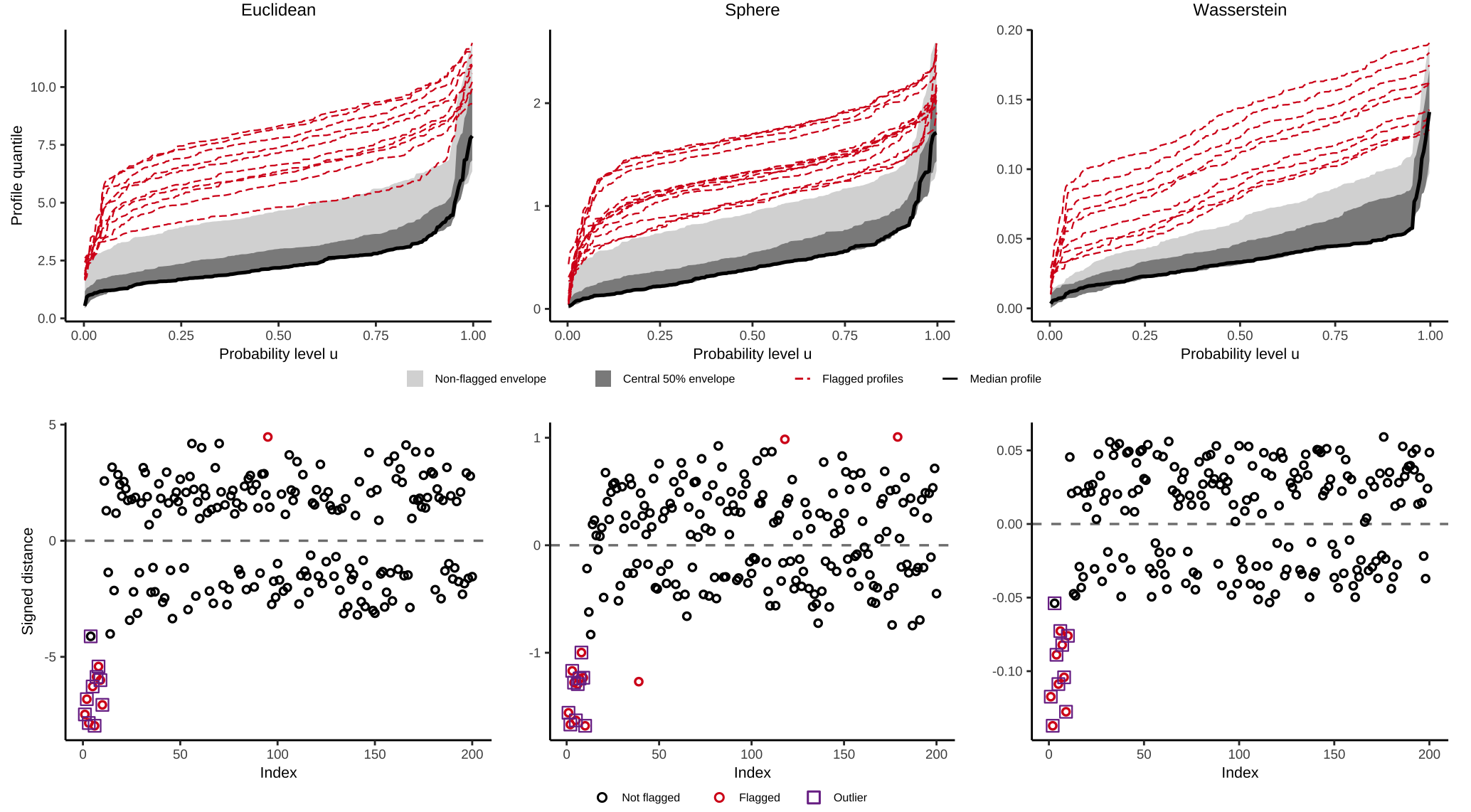}
\caption{Outlier detection for 200 objects, including 10 outliers, in the Euclidean space of Example~\ref{exmp::1} (left), the sphere of Example~\ref{exmp::2} (center), and the Wasserstein space of Example~\ref{exmp::8} (right). The first row presents distance-profile-based boxplots, with detected outliers highlighted in red. The second row displays signed distance plots, with detected outliers marked by red circles and the true outliers by squares. The proposed procedure correctly identifies 9, 10, and 9 of the ten true outliers, respectively.}\label{fig::boxplot}
\end{figure}

Observe that profiles are distributions on the half-line, we adapt functional boxplot techniques \citep{HS10,sun2011functional,DG18} to visualize object-valued residuals. Specifically, we construct a functional boxplot in the distance-profile domain using the quantile functions $u\mapsto\widehat G_i^{-1}(u)$, with the center-outward ordering determined by the proposed scores $\{\widehat C_i\}_{i=1}^n$. The display superimposes four elements, each plotted against $u\in(0,1)$:
\begin{itemize}
\item \textbf{Median.} The profile $\widehat G_{[1]}^{-1}$ of the medoid $\mathcal E_{[1]}$, which plays the role of the median line;
\item \textbf{Box.} The envelope of the profiles of the central half, which plays the role of the box as $\left[\min_{i\in\widehat{\mathcal C}_{1/2}}\widehat G_i^{-1}(u),\ \max_{i\in\widehat{\mathcal C}_{1/2}}\widehat G_i^{-1}(u)\right]$, $u\in(0,1)$;
\item \textbf{Whiskers.} The envelope of the profiles of all residuals that are not flagged by the rule below, which plays the role of the whiskers;
\item \textbf{Outliers.} The individually drawn profiles of the residuals flagged under the rule below;
\end{itemize}
A profile that lies above the box at every level $u$ belongs to a residual that is farther from most of the sample than the central residuals are. The display thus shows how far, and at which range of distances, a flagged residual departs from the bulk. Figure~\ref{fig::boxplot} shows the display for the three spaces of Examples~\ref{exmp::1}, \ref{exmp::2} and~\ref{exmp::8} using the outlier detection rule introduced later.

A large mean-distance score $\Delta(z)$, or equivalently a small centrality score $C(z)$, indicates an outlying object. Outlying residuals therefore lie in the upper tail of the empirical scores $\widehat\Delta_1,\ldots,\widehat\Delta_n$.  We flag a residual whose score exceeds an upper fence in the sense of \citet{tukey1977exploratory}, constructed from the quartiles of these scores. Recall that their empirical $1/4$- and $3/4$-quantiles are $\widehat q_{1/4}=\widehat\Delta_{[\lceil n/4\rceil]}$ and $\widehat q_{3/4}=\widehat\Delta_{[\lceil 3n/4\rceil]}$, respectively. For a fixed constant $\kappa>0$, the fence and the set of flagged residuals are
\begin{align}\label{eq::fence}
\widehat\tau=\widehat q_{3/4}+\kappa\left(\widehat q_{3/4}-\widehat q_{1/4}\right),\qquad
\widehat{\mathcal O}=\{ \mathcal E_i \in \{\mathcal E_i\}_{i=1}^n \mid \widehat\Delta_i>\widehat\tau\}.
\end{align}
Their population counterparts are the fence $\tau=q_{3/4}+\kappa(q_{3/4}-q_{1/4})$ and the outlier region $\mathcal O=\{z\in\Omega \mid \Delta(z)>\tau\}$. Following the classical boxplot, we set $\kappa=3/2$ by default.

\begin{remark}
\citet{liu2026wasserstein} flag a density when its centrality score falls below the median centrality score by more than three times the range of the largest half of the centrality scores. 
In the present notation, their rule is the upper fence $\widehat q_{1/2}+3(\widehat q_{1/2}-\widehat\Delta_{[1]})$ on the mean distances. 
The range $\widehat q_{1/2}-\widehat\Delta_{[1]}$ is the spread of the mean-distance function over the central half of the sample. 
The mean-distance function is the expectation of a distance that is flat near its minimum, so this range is small relative to the spread of the scores in the outer half, and its size depends on the geometry of the space. 
For a standard Gaussian population in $\mathbb R^p$, their rule flags about 15\% of the population when $p=1$ and none of it when $p=10$, whereas the fence~\eqref{eq::fence} with $\kappa=3/2$ flags about 6\% when $p=1$ and about 1\% when $p=10$. 
Our fence measures the spread of the scores where they are informative, between the first and third quartiles. 
A skewness adjustment of the constant $\kappa$, as in \citet{hubert2008adjusted}, can be applied to it in the same way as \citet{liu2026wasserstein} apply one to their fence.
\end{remark}

\subsection{Residual-based prediction set}\label{sect::PI}

The ordering techniques in Section~\ref{sect::boxplot} can also be used to define a predictive region. For a nominal level $\alpha\in(0,1)$, let $\widehat{q}_{1-\alpha}={\rm Quantile}_{1-\alpha}(\widehat\Delta_1,\widehat\Delta_2,\ldots,\widehat\Delta_n)$ be the empirical $(1-\alpha)$-quantile of the scores of Section~\ref{sect::boxplot}. For a given future predictor $X_{n+1}$ and candidate response $y\in\Omega$, write $\mathcal{E}(y)=T_{r(X_{n+1}),y}(\mu_Y)$. The proposed $(1-\alpha)$-prediction set is 
\begin{align}\label{formula::PI}
\widehat{\mathcal{P}}_{1-\alpha}(X_{n+1}) = \left\{ y\in\Omega : n^{-1}\sum_{i=1}^n d(\mathcal{E}(y),\mathcal E_i) \leq \widehat{q}_{1-\alpha} \right\}.    
\end{align}
The next theorem establishes the asymptotic validity of the proposed prediction set.

\begin{theorem}\label{thm::PI}
Let $F_\Delta(t)=\mathbb P\{\Delta(\mathcal E)\leq t\}$ and $q_{1-\alpha}=F_\Delta^{-1}(1-\alpha)$. If $F_\Delta$ is continuous at $q_{1-\alpha}$ and $F_\Delta(q_{1-\alpha}-\xi)<1-\alpha<F_\Delta(q_{1-\alpha}+\xi)$ for every $\xi>0$, then $\mathbb P\{Y_{n+1}\in\widehat{\mathcal P}_{1-\alpha}(X_{n+1})\}=1-\alpha+o(1)$.
\end{theorem}

\begin{remark}
Sufficient conditions for transferring the conclusions of Theorem~\ref{thm::PI} to the fitted residual setting are $n^{-1}\sum_{i=1}^n d(\widehat{\mathcal E}_i,\mathcal E_i)=o_p(1)$ and $d(\widehat{\mathcal E}(Y_{n+1}),\mathcal E(Y_{n+1}))=o_p(1)$ where $\widehat{\mathcal E}(Y_{n+1})=T_{\widehat r(X_{n+1}),Y_{n+1}}(\widehat\mu_Y)$.
\end{remark}

\section{Simulation}\label{sect::simulation}

We examine the proposed methodologies in the Euclidean, spherical, and Wasserstein spaces of Examples~\ref{exmp::1}, \ref{exmp::2} and \ref{exmp::8}. All reported simulation summaries use 1000 Monte Carlo replications. Here we focus on the verification of the conditional centering property \eqref{formula::conditional centering} in Definition~\ref{def::metric subtraction}. Additional simulation studies on the conditional Fr\'echet mean test in Section~\ref{sect::residual test} and the residual-based prediction set in Section~\ref{sect::PI} can be found in Sections~S5.1 and S5.2 of the supplement, respectively.


For each replication, we generate $n$ independent responses with known population Fr\'echet mean $\mu_Y$, choose a fixed reference object $\omega$, and construct $\mathcal E_i=T_{\mu_Y,Y_i}(\omega)$. Let $\widehat\mu_{\mathcal E}$ be the sample Fr\'echet mean of the residuals. We record the squared error $d^2(\widehat\mu_{\mathcal E},\omega)$ and report its average across replications as the mean squared error (MSE). If the proposed metric subtraction operator satisfies the conditional centering property \eqref{formula::conditional centering} in Definition~\ref{def::metric subtraction}, the MSE is expected to converge to zero as the sample size $n$ increases. Here, we consider the sample size $n\in\{100,200,\dots,1000\}$.

For the Euclidean space in Example~\ref{exmp::1}, the observations are i.i.d.~generated from the Gaussian distribution $Y_i\sim N(\mu_Y,\Sigma)$ where $\mu_Y=(1,-1,0.5,2,-0.5)^{\top}$ and $\Sigma$ is the $5\times5$ covariance matrix with diagonal entries equal to $1$ and off-diagonal entries equal to $0.3$. The reference point is chosen as $\omega=(-1,0.5,1,-0.5,2)^{\top}$.

For the sphere in Example~\ref{exmp::2}, the observations are i.i.d.~generated from the von Mises--Fisher distribution with $\mu_Y=(1,0,0)^\top$ and concentration parameter~10. The reference point is $\omega=(1/\sqrt{2},1/\sqrt{2},0)^\top$.

For the Wasserstein space in Example~\ref{exmp::8}, the true Fr\'echet mean is chosen to be the uniform distribution on $[0,1]$, with quantile function $Q_{\mu_Y}(u)=u$, $0\leq u\leq1$. We generate random quantile functions via $Q_i(u)=u+A_i+B_i\sin(2\pi u)$ where $A_i\stackrel{\rm iid}{\sim}{\rm Unif}(-0.05,0.05)$ and $B_i\stackrel{\rm iid}{\sim}{\rm Unif}(-0.04,0.04)$. This ensures that each $Q_i$ is a valid strictly increasing quantile function. The reference point is chosen to have quantile function $Q_\omega(u)=0.15+0.7u$ corresponding to a uniform distribution on $[0.15,0.85]$. 

\begin{figure}[!htb]
\centering
\includegraphics[width=\linewidth]{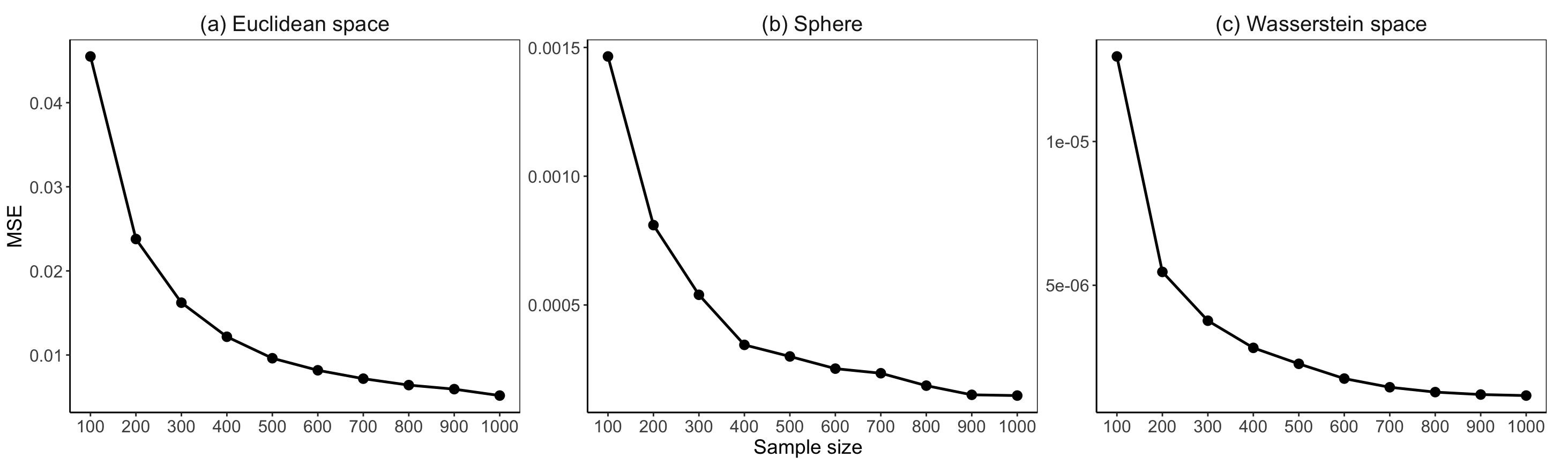}
\caption{MSE of the sample Fr\'echet mean of the residuals for (a) the Euclidean space, (b) the sphere, and (c) the Wasserstein space.}\label{fig::sim subtraction}
\end{figure}

Figure~\ref{fig::sim subtraction} reports that the MSE of the sample Fr\'echet mean $\widehat\mu_{\mathcal E}$ decreases as the sample size increases, supporting Proposition~S1 that the conditional centering property \eqref{formula::conditional centering} in Definition~\ref{def::metric subtraction} holds.

\section{Empirical data analysis}\label{sect::applications}

\subsection{New York City Citi Bike trip volume shapes}\label{sect::NYC bike}

Historical trip records from the New York City Citi Bike program are available at \url{https://citibikenyc.com/system-data}. We analyze the daily shape of trip activity across the day, focusing on the $n=365$ days of 2019. For each day, trip counts are aggregated into 30-minute intervals and represented at the interval midpoints. Normalizing the resulting daily profile to integrate to one produces a probability density, so the response describes the shape of within-day usage rather than the total number of trips. This data set corresponds to the Wasserstein space in Example~\ref{exmp::8}.

\begin{figure}[!htb]
\centering
\includegraphics[width=\linewidth]{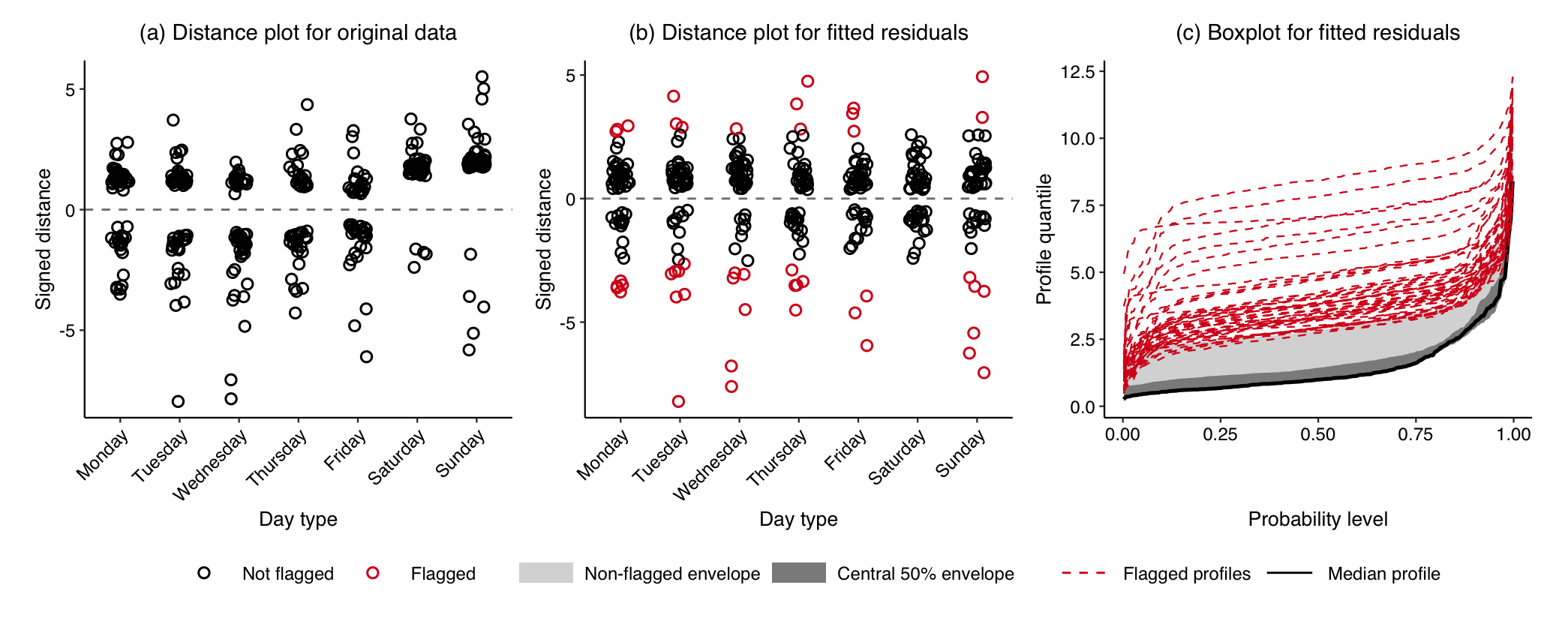}
\caption{Diagnostics for daily Citi Bike trip volume shapes in 2019. (a) Signed distance plot of the observed distributions by day type, (b) signed distance plot of the fitted residuals, (c) boxplot of the fitted residuals. Red circles in (b) and red dashed profiles in (c) identify residuals flagged by the mean-distance fence.}\label{fig::NYC bike 1}
\end{figure}

The predictor is day type, with categories $\{{\rm Monday},{\rm Tuesday},\ldots,{\rm Sunday}\}$. Since the predictor is categorical, signed distances can be visualized directly across day types. Figure~\ref{fig::NYC bike 1}(a) displays the signed distance plot of the original bicycle trip volume shapes, revealing systematic differences between weekdays and weekends. In particular, the plot suggests that the conditional Fr\'echet means for weekends differ from those for weekdays. To formally assess these differences, we apply the constant conditional Fr\'echet mean test of Section~\ref{sect::residual test} directly to the response objects. The test rejects the null hypothesis in \eqref{formula::null hypothesis}, providing statistical evidence that the conditional Fr\'echet mean varies across day types.

We fit a category-specific Fr\'echet mean model using the procedure described in Section~S3 and construct distribution-valued residuals through metric subtraction. Figure~\ref{fig::NYC bike 1}(b) displays the signed distance plot of the fitted residuals. In contrast to the original observations in Figure~\ref{fig::NYC bike 1}(a), the residuals exhibit no clear systematic differences in conditional Fr\'echet means across day types, suggesting that the fitted model adequately captures the category-specific location effects. We next evaluate whether the fitted model is underfitted based on the constant conditional Fr\'echet mean test. Here, we consider the season as an additional categorical predictor. We apply the test to the fitted residuals with season as the predictor, so that the null hypothesis states that the Fr\'echet mean conditional on the season is a constant. The test rejects the null hypothesis, indicating that the fitted model is underfitted and that seasonal information should be included in the model.

The boxplot in Figure~\ref{fig::NYC bike 1}(c) flags 47 residuals as outliers, highlighted in red in panel (b). Their large mean-distance scores indicate atypical usage shapes relative to the fitted day-type model. Such observations may represent unusual daily patterns and merit inspection rather than automatic exclusion. Notably, these 47 outliers include six observations corresponding to New Year's Day, Memorial Day, Independence Day, Labor Day, Thanksgiving Day, and Christmas Day, which are treated as weekdays in the original data set.

\begin{figure}[!htb]
\centering
\includegraphics[width=\linewidth]{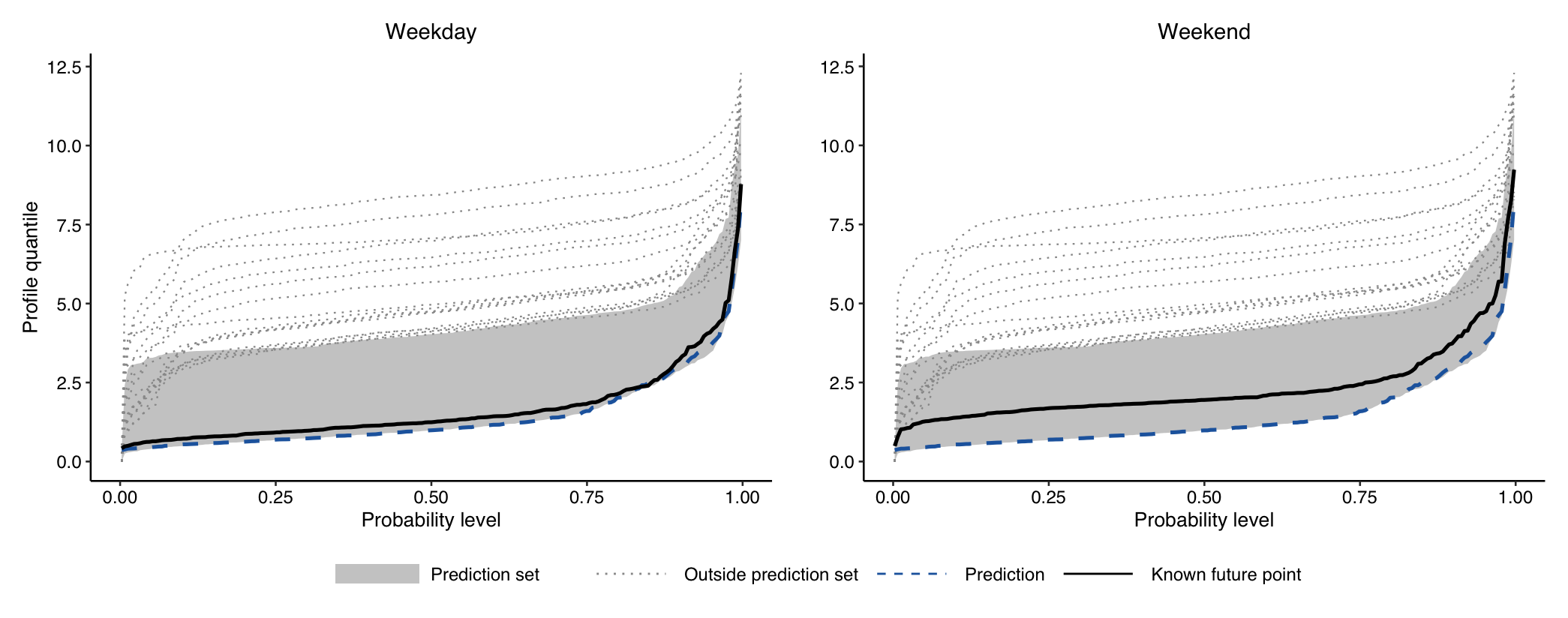}
\caption{Boxplots for the prediction sets for a weekday future point (left), and a weekend future point (right). Gray shading represents the prediction region in the profile display, the blue dashed curve represents the category-specific point prediction, and the black solid curve represents the observed future response.}\label{fig::NYC bike PI}
\end{figure}

We then illustrate the residual-based prediction sets developed in Section~\ref{sect::PI}. For one weekday type and one weekend type, we randomly select one observation as a future response, refit the model using the remaining 364 observations, and construct a prediction set for the held-out response. Figure~\ref{fig::NYC bike PI} displays the resulting 95\% prediction sets for the selected weekday and weekend observations. In both cases, the held-out response is contained in its corresponding prediction set.

Section~S6.1 of the supplement examines a refit after removing the 47 outliers. In the reported category-stratified ten-fold cross-validation, the refitted model has a smaller mean squared prediction error of 1.61, compared with 3.25 for the fit using all observations. Moreover, following the residual test results, we consider another model with predictors containing both day type and season information, resulting in 28 day-type–season combinations.
\begin{figure}[!htb]
\centering
\includegraphics[width=\linewidth]{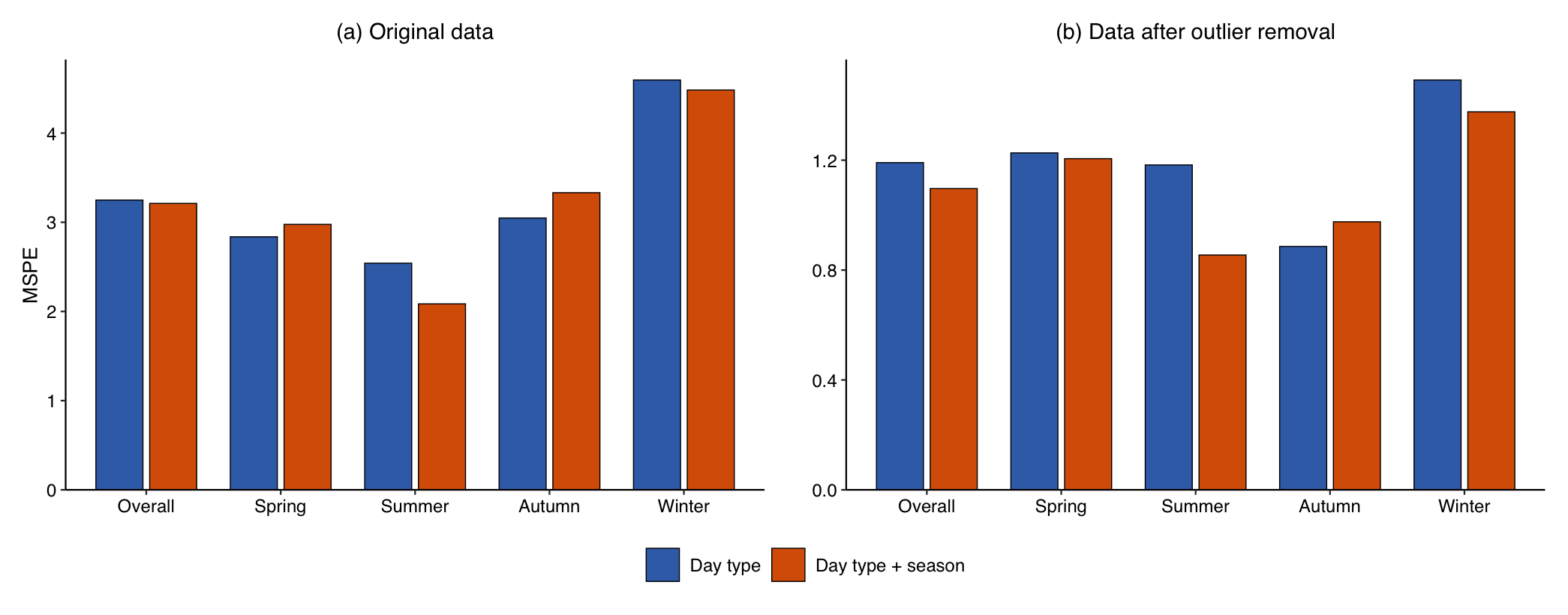}
\caption{Mean squared prediction errors (MSPEs) for (a) original daily Citi Bike trip volume shapes, and (b) data after outlier removal, comparing the model with day types as the predictor (blue) and the model with day-type–season combinations as the predictor (orange).}\label{fig::NYC bike prediction compare}
\end{figure}

Figure~\ref{fig::NYC bike prediction compare} presents the average mean squared prediction errors, calculated via the squared 2-Wasserstein distances between the predicted point and the true future point after ten-fold cross-validation. For the original data set, incorporating seasonal information yields a modest improvement in predictive accuracy, with the largest reduction in prediction error observed for summer observations. Following outlier removal, both models produce lower prediction errors, while the advantage of incorporating seasonal information becomes more pronounced. These findings suggest that accounting for seasonal variation improves predictive performance, particularly after the influence of atypical observations has been reduced.

\subsection{U.S. electricity generation compositions}

The electricity data are monthly net generation measurements obtained from \url{https://www.eia.gov/electricity/data/browser/}. Following \citet{xu2025quantifying}, we convert generation from seven fuel types, coal, petroleum, gas, nuclear, conventional hydroelectric, renewables, and solar, to relative shares of total generation. The analysis uses $n=120$ observations from January 2001 through December 2010. Each response lies in the simplex $\Delta^6=\{(\delta_1,\ldots,\delta_7)^\top\in\mathbb R^7:\sum_{\ell=1}^7\delta_\ell=1,\;\delta_\ell\geq0\}$. As discussed in Example~\ref{exmp::6}, for two compositions $x,y\in\Delta^6$, we use the Fisher--Rao distance.

\begin{figure}[!htb]
\centering
\includegraphics[width=\linewidth]{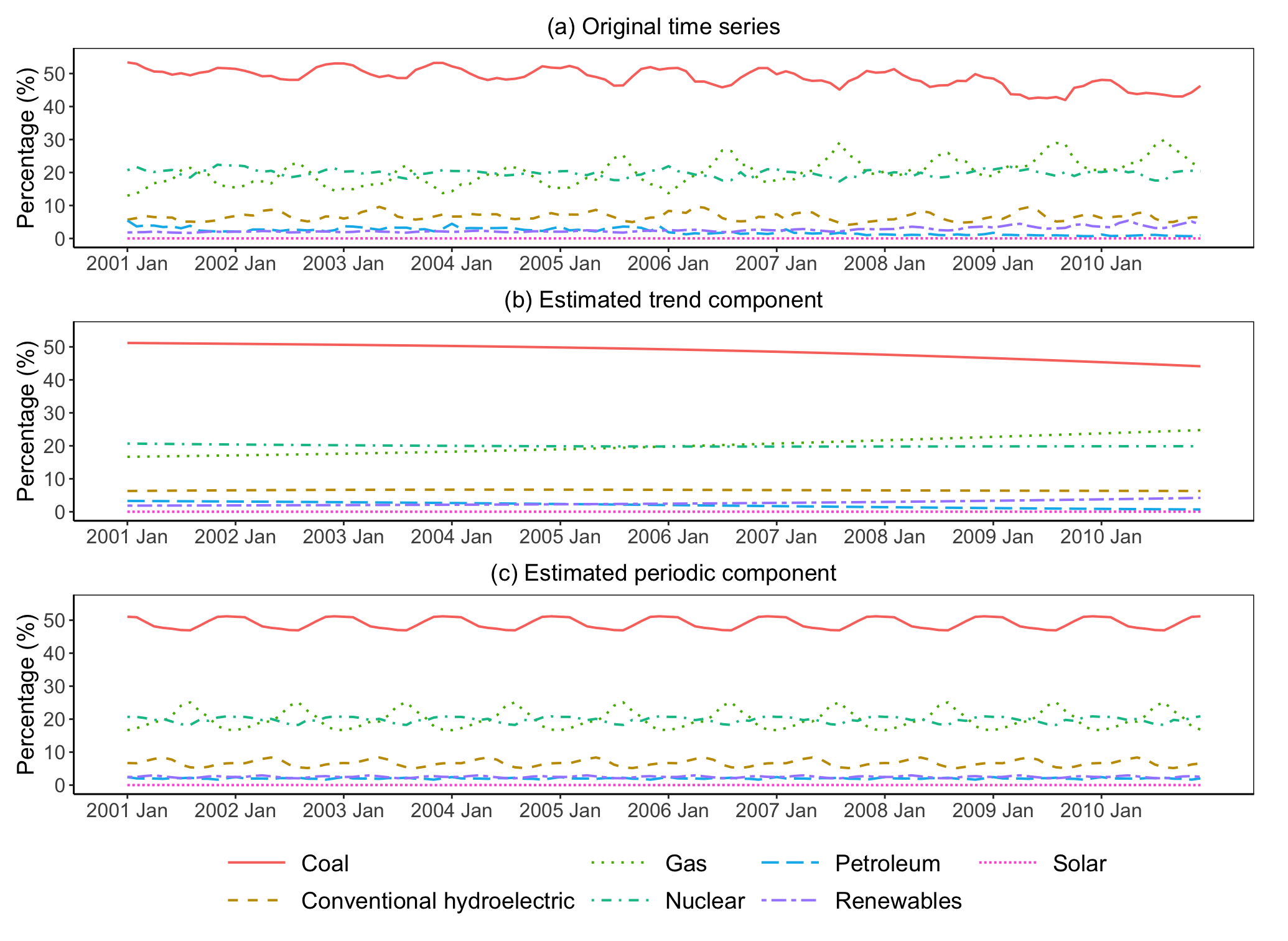}
\caption{Time plots of (a) original electricity generation compositions, (b) estimated trend component, and (c) estimated periodic component. Curves correspond to the seven fuel types.}\label{fig::US enengry raw}
\end{figure}

\begin{figure}[!htb]
\centering
\includegraphics[width=\linewidth]{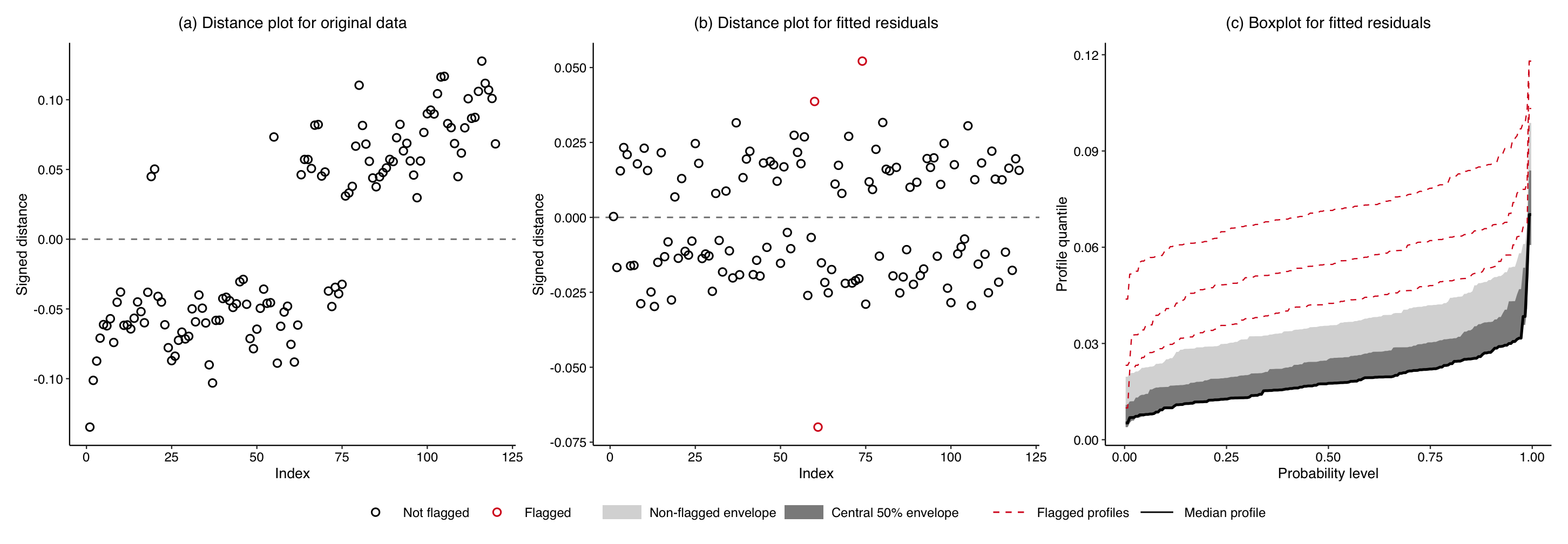}
\caption{Diagnostics for U.S. electricity generation compositions. (a) Signed distance plot of the original electricity generation compositions, (b) signed distance plot of the fitted residuals after removing trend, periodicity, and autoregressive structure, (c) boxplot of the fitted residuals. Red circles in (b) and red dashed profiles in (c) identify residuals flagged by the mean-distance fence.}\label{fig::US enengry 1}
\end{figure}

Figure~\ref{fig::US enengry raw}(a) displays the original electricity generation compositions, which exhibit both trend and periodicity. The signed distance plot in Figure~\ref{fig::US enengry 1}(a) further reveals these temporal patterns, indicating departures from stationarity driven by trend and periodic components. These findings are consistent with the patterns observed in Figure~\ref{fig::US enengry raw}. Notably, although the underlying trend is not readily discernible from either the original compositional time series in Figure~\ref{fig::US enengry raw}(a) or the estimated trend component in Figure~\ref{fig::US enengry raw}(b), it becomes substantially more apparent in the signed distance plot in Figure~\ref{fig::US enengry 1}(a). This highlights the utility of the proposed visualization in revealing systematic temporal structures that may otherwise be difficult to identify directly from object-valued observations.

We apply the trend-periodic decomposition of \citet{xu2025quantifying} sequentially. First, we estimate the trend component, shown in Figure~\ref{fig::US enengry raw}(b), and remove it using metric subtraction. We then estimate the periodic component from the detrended series, shown in Figure~\ref{fig::US enengry raw}(c), and remove it in the same way. Next, we fit a spherical autoregressive model \citep{zhu2024spherical} to the resulting series and obtain the final compositional residuals by removing the fitted autoregressive component through metric subtraction. The signed distance plot of these residuals, displayed in Figure~\ref{fig::US enengry 1}(b), exhibits no discernible systematic temporal pattern, suggesting that the conditional Fr\'echet mean of the residuals is constant over time. To formally assess this property, we apply the test developed in Section~\ref{sect::residual test}, taking the normalized time index on $[0,1]$ as the predictor $X$. The test fails to reject the null hypothesis in \eqref{formula::null hypothesis}, consistent with the graphical evidence in Figure~\ref{fig::US enengry 1}(b).

The signed distance plot in Figure~\ref{fig::US enengry 1}(b) also suggests the presence of potential outliers among the fitted residuals. To investigate these observations, we apply the distance-profile-based boxplot developed in Section~\ref{sect::boxplot}, with the results displayed in Figure~\ref{fig::US enengry 1}(c). Three observations, corresponding to December 2005, January 2006, and February 2007 with indices 60, 61, and 74, respectively, are identified as outliers and highlighted by red circles in Figure~\ref{fig::US enengry 1}(b). These observations exhibit relatively large distances from the reference point in the signed distance plot, while their distance profiles deviate substantially from the central profiles in the boxplot. The agreement between the two visualizations illustrates the utility of the proposed diagnostic tools in identifying atypical object-valued residuals. Further analysis after removing the identified outliers is provided in Section~S6.2 of the supplementary material.

\begin{figure}[!htb]
\centering
\includegraphics[width=\linewidth]{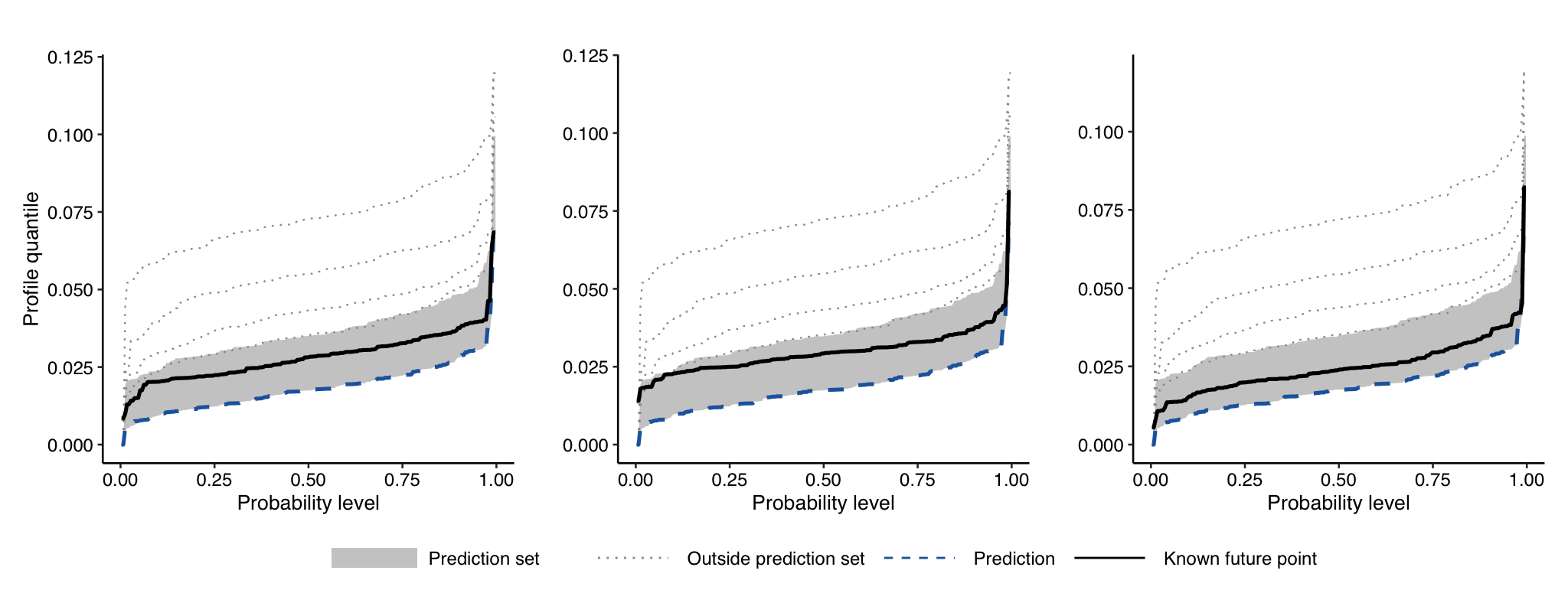}
\caption{Boxplots for the prediction sets produced under three expanding windows, with each successive window incorporating one additional observation to predict the next response. Gray shading represents the prediction region in the profile display, the blue dashed curve represents the one-step-ahead point prediction, and the black solid curve represents the observed future response.}\label{fig::US enengry PI}
\end{figure}

We then illustrate the residual-based prediction set developed in Section~\ref{sect::PI}. We fit the model using the expanding window $\{Y_1,\ldots,Y_{116+k}\}$ and predict $Y_{117+k}$ for $k\in\{1,2,3\}$, that is, we predict the compositions for the final three months of 2010. Figure~\ref{fig::US enengry PI} displays the corresponding 95\% prediction sets through their distance profiles, together with the point predictions and observed future responses. All three future responses are contained in their respective sets. 

\section{Conclusion}\label{sect::discussion}

In this paper, we develop a metric subtraction framework that constructs object-valued residuals in the original response space by transporting geodesic to a common reference object. Central to this framework is the conditional centering property, which requires that subtracting the true conditional Fr\'echet mean yields residuals with a constant conditional Fr\'echet mean. We establish sufficient conditions that guarantee this property.

Building on the proposed metric subtraction framework, we develop a suite of diagnostic and inferential tools for object-valued residuals. Specifically, signed distance plots facilitate the identification of systematic patterns, a randomized test assesses the constancy of the conditional Fr\'echet mean, distance-profile-based boxplots characterize centrality and outlyingness, and residual-based prediction sets quantify uncertainty for future random objects.

Several directions merit further investigation. One is to extend the proposed testing framework to more general conditional hypotheses, potentially through another multiplier-assisted construction using sample splitting in \citet{cai2026match}. More broadly, the proposed metric subtraction framework opens avenues for statistical analysis of object-valued residuals beyond the i.i.d. setting. In particular, it provides a foundation for investigating temporal dependence through stationarity testing \citep{zhu2026stationarity} and long-memory estimation \citep{seo2026quantifying}.


\section*{Supplementary Material}
Verification for Examples~\ref{exmp::1}--\ref{exmp::8} (Section~S1). Properties of the Nadaraya--Watson Fr\'echet regression estimator used in the residual test (Section~S2), implementation of the residual test for discrete predictors (Section~S3), proofs of the main results (Section~S4), additional simulations (Section~S5), and additional real data analyses (Section~S6).

\if0\anon
{
} \fi

\if1\anon
{
\section*{Acknowledgements}
This research was supported by an Australian Research Council Future Fellowship (FT240100338) and undertaken with assistance from computational resources provided by the Australian Government through the National Computational Infrastructure (NCI) under the Macquarie University Merit Allocation Scheme.
} 
\fi

\textbf{Data availability statement:} Historical New York City Citi Bike trip records are publicly available at \url{https://citibikenyc.com/system-data}. The U.S. electricity generation data are publicly available from the U.S. Energy Information Administration at \url{https://www.eia.gov/electricity/data/browser/}. The computational code is available at \url{https://github.com/d5rhtkp4kc-alt/metric-subtraction/}.

\textbf{Statement on AI use:} We used AI tools (ChatGPT 5.6 Sol, Claude Opus 5, and Claude Fable 5.1) for language editing, and for programming code, we used them for minor syntax checking, visual alignment, layout adjustments, and debugging. All outputs were thoroughly verified and finalized by the authors. All scientific results are our own.

\textbf{Disclosure statement:} The authors declare no relevant financial or non-financial competing interests.  



\begin{spacing}{1.6}
\setlength{\bibsep}{2pt}
\bibliography{reference}
\end{spacing}

\end{document}